\documentclass[%
 reprint,%reprint,
superscriptaddress,
 amsmath,amssymb,
prb,
]{revtex4-1}

\usepackage{graphicx}% Include figure files
\usepackage{dcolumn}% Align table columns on decimal pointuse
\usepackage{color}
\usepackage{ulem}
\usepackage{float}
\usepackage{hyperref}
\usepackage{nameref}
\begin{document}

%\title{Type-I to type-II band alignment switching for (In,Ga)(As,Sb)/GaAs/GaP quantum dots overgrown by a thin GaSb capping layer}
%
%\title{Anomalous temperature dependence of (In,Ga)(As,Sb)/GaAs/GaP quantum dots overgrown by a thin GaSb capping layer revealing Type-I to type-II band alignment switching}
%
%\title{Anomalous temperature dependence of (In,Ga)(As,Sb)/GaAs/GaP quantum dots overgrown by a thin GaSb capping layer}
%
%\title{Anomalous temperature dependence associated with ${\bf k}$-momentum and spatially indirect transitions in (In,Ga)(As,Sb)/GaAs/GaP quantum dots overgrown by a thin GaSb capping layer}
%
%\title{Anomalous temperature dependence of (In,Ga)(As,Sb)/GaAs/GaP quantum dots overgrown by a thin GaSb capping layer associated with ${\bf k}$-momentum and spatially indirect excitons}
%
%\title{Anomalous temperature dependence associated with momentum and spatially indirect excitons in (In,Ga)(As,Sb)/GaAs/GaP quantum dots overgrown by a thin GaSb capping layer}
%
%\title{Anomalous temperature dependence of excitons in (In,Ga)(As,Sb)/GaAs/GaP quantum dots overgrown by a thin GaSb capping layer for nanomemory applications}
%
\title{Anomalous luminescence temperature dependence of (In,Ga)(As,Sb)/GaAs/GaP quantum dots overgrown by a thin GaSb capping layer for nanomemory applications}
%
%\author{Elisa Maddalena Sala\orcidlink{0000-0001-8116-8830}} 
\author{Elisa Maddalena Sala} 
\email[]{e.m.sala@sheffield.ac.uk}
\affiliation{Center for Nanophotonics, Institute for Solid State Physics, Technische Universit\"{a}t Berlin, Germany}
%\affiliation{Technische Universit\"{a}t Berlin, Institut f\"{u}r Festk\"{o}rperphysik, Hardenbergstr. 36, 10623 Berlin, Germany}
\affiliation{EPSRC National Epitaxy Facility, The University of Sheffield, North Campus, Broad Lane, S3 7HQ Sheffield, United Kingdom}
\affiliation{Department of Electronic and Electrical Engineering, The University of Sheffield, North Campus, Broad Lane, S3 7HQ Sheffield, United Kingdom}

%\author{Petr Klenovsk\'y\orcidlink{0000-0003-1914-164X}}
\author{Petr Klenovsk\'y}
\email[]{klenovsky@physics.muni.cz}
\affiliation{Department of Condensed Matter Physics, Faculty of Science, Masaryk University, Kotl\'a\v{r}sk\'a~267/2, 61137~Brno, Czech~Republic}
%\affiliation{Central European Institute of Technology, Masaryk University, Kamenice 753/5, 62500~Brno, Czech~Republic}
\affiliation{Czech Metrology Institute, Okru\v{z}n\'i 31, 63800~Brno, Czech~Republic}

\date{\today}

\begin{abstract}
We study (In,Ga)(As,Sb)/GaAs quantum dots embedded in a GaP (100) matrix, which are overgrown by a thin GaSb capping layer with variable thickness. Quantum dot samples are studied by temperature-dependent photoluminescence, and we observe that the quantum dot emission shows anomalous temperature dependence,~i.e., increase of energy with temperature increase from 10~K to $\sim$70~K, followed by energy decrease for larger temperatures. With the help of fitting of luminescence spectra by Gaussian bands with energies extracted from eight band ${\bf k}\cdot{\bf p}$ theory with multiparticle corrections calculated using the configuration interaction method, we explain the anomalous temperature dependence as mixing of momentum direct and indirect exciton states. We also find that the ${\bf k}$-indirect electron-hole transition in type-I regime at temperatures $<70$~K is optically more intense than ${\bf k}$-direct. Furthermore, we identify a band alignment change from type-I to type-II for QDs overgrown by more than one monolayer of GaSb. Finally, we predict the retention time of (In,Ga)(As,Sb)/GaAs/AlP/GaP quantum dots capped with GaSb layers with varying thickness, for usage as storage units in the QD-Flash nanomemory concept and observe that by using only a 2~ML-thick GaSb capping layer, the projected storage time surpasses the non-volatility limit of 10 years.

\end{abstract}

%\pacs{73.21.La, 75.75.-c, 85.35.Be, 68.65.Hb}
%
\vspace{0.5cm}

\maketitle

%\begin{flushleft}
%\textbf{TABLE OF CONTENTS}
%\end{flushleft}

%\renewcommand*\contentsname{\small{TABLE OF CONTENTS}}

%\tableofcontents

%\setcounter{tocdepth}{3}
\setcounter{secnumdepth}{3}

\section{Introduction}
\label{sec:intro}

Self-assembled Quantum Dots (QDs) from III-V semiconductor compounds have been extensively studied in the last decades, thanks to their distinctive physical properties. They have been employed in telecommunication devices such as lasers and amplifiers~\cite{Bimberg1997,Ledentsov2010,Ledentsov, Heinrichsdorff1997, Schmeckebier2017, Unrau_laserphotonics_2014}, as single and entangled photon emitters for quantum communication technologies~\cite{michler_quantum_2017, muller_quantum_2018, salter_entangled-light-emitting_2010,Krapek2010,PetrPRL2021}, and as building blocks for nanomemory devices known as ``QD-Flash"~\cite{Marent2009_microelectronics, Marent_APL2007_10y, Stracke2014, Bonato2016_PSSB, Sala2018, t_sala}. 

Particularly for the latter application, QDs showing a type-II band alignment~\cite{Klenovsky2010, Klenovsky_IOP2010, Klenovsky2015, Klenovsky2016, Hayne_2013} are required for maximizing the hole localization energy, and in turn to extend their storage time in memory cells~\cite{Bimberg2011_SbQDFlash, Marent_APL2007_10y, Bonato2016_PSSB, Sala2018, t_sala}. Promising candidates are Sb-based QDs embedded in GaP or AlP matrix materials. In this respect, GaP has the advantage of allowing for opto-electronic device integration on Si due to its lattice-mismatch as low as 0.4\,\%~\cite{Grassman_apl2013, Beyer_jap2013, MOVPE_GaP_Si}. Thus, GaP is considered to be a promising matrix material for the growth of such QDs and related nanomemory devices. Recently, efforts have been directed in the fabrication of Sb-based QDs embedded in GaP and improvements in the retention time for QDs at room temperature have been obtained. In particular, the record storage time at room temperature to date is of 4~days for Molecular Beam Epitaxy (MBE)-grown GaSb/GaP QDs by Bonato~\textit{et al.}~\cite{Bonato2016_PSSB}. For Metal-Organic Vapor Phase Epitaxy (MOVPE)-grown In$_{0.5}$Ga$_{0.5}$As/GaAs/AlP QDs in a GaP matrix~\cite{Bonato_APL2015, Stracke2014}, it was found that introducing Sb during growth led to the improvement of one order of magnitude in the storage time, reaching about 1~hour for (In,Ga)(As,Sb)/GaAs/AlP/GaP QDs by Sala~\textit{et al.}~\cite{Sala2018, t_sala}. This result represents the storage time record for MOVPE-grown QDs so far. Moreover, employing MOVPE, instead of MBE, to fabricate such Sb-based QDs will allow for a cost-effective and large-scale fabrication of QDs and related nanomemory devices. Following the aforementioned result, detailed morphological investigations by means of Cross-Sectional Tunneling Microscopy (XSTM) combined with Atom Probe Tomography (APT) confirmed the incorporation of Sb into the QDs to an extent of 10-15$\%$~\cite{RajaNature}. However, upon detailed optical and theoretical analysis~\cite{Steindl_2019, Steindl_2021}, we found that such QDs still present a type-I band alignment. 

A growth strategy usually employed to change type-I QDs into type-II is the overgrowth of QDs by Sb-based capping layer immediately after the QD formation, which maximizes the Sb incorporation into the QD layer. Examples of this approach are found for type-I InAs/GaAs QDs, which change into type-II upon overgrowth with Ga(As,Sb) layers~\cite{Jin2007,Klenovsky2010}. We also previously considered capping our (In,Ga)(As,Sb)/GaAs/GaP QDs with an Sb-containing layer, namely GaSb, with a thickness of ca. 1~ML~\cite{Steindl_2019, Steindl_2021}. There, we found that introducing the GaSb cap effectively modified the QD composition thereby increasing the Sb content, and leading to an energy swapping of $\Gamma$ (1.732 eV) and $L$ (1.701 eV) states compared to the initial values of 1.725~eV and 1.755~eV, respectively, where also an increased leakage of the electron wave function outside the QD body was induced~\cite{Steindl_2019, Steindl_2021}.

\setcounter{tocdepth}{1}

In this work, we take the next step and study more in detail the effect of a thickness variation of a GaSb capping layer overgrowing the (In,Ga)(As,Sb)/GaAs/GaP QDs. We study the optical properties of the QDs overgrown with the GaSb layer of thickness ranging from 0 (i.e. no cap) to 1.4~ML. We perform temperature-dependent photoluminescence~(PL) investigations and compare the obtained results with theoretical calculations using eight-band $\mathbf{k\cdot p}$~\cite{Birner:07,Mittelstadt2022} method with multiparticle corrections computed by the configuration interaction (CI) algorithm.~\cite{Klenovsky2015,Klenovsky_PRB2018} 

The PL spectra exhibit an anomalous temperature dependence of QD emission, which we interpret using detailed fitting with exciton emission energies taken from our theory and reproduce the anomalous temperature shift with simple model. We also find that in our system for type-I confinement at low temperatures, the dominant transition is between L electrons and $\Gamma$ holes, while with temperature increase the $\Gamma$-$\Gamma$ transition becomes more prominent.

Furthermore, we show that, by increasing the GaSb thickness, our QDs can be turned from type-I into type-II confinement. We thus present a method to effectively obtain type-II QDs with increased hole localization, for extending the storage time in QD-Flash devices. For example, we predict that using a GaSb capping layer with thickness greater than 2~ML will lead to an increase of the storage time of (In,Ga)(As,Sb)/GaAs/AlP/GaP QDs by more than 10 years, thus surpassing the non-volatility limit. 

The manuscript is organized as follows. After the introduction in Sec.~\ref{sec:intro}, we describe our samples in Sec.~\ref{sec:Fab} and PL results in Sec.~\ref{sec:PL}. The PL spectra are analysed using theory given in Sec.~\ref{sec:calcs} and interpreted in Sec.~\ref{sec:fitting}. Thereafter, in Sec.~\ref{sec:qdflash} we give the prediction of QD Flash storage times and finally conclude in Sec.~\ref{sec:conclusions}.

%%%%%%%%%%%%%%%%%%%%%%%%%%%%%%%%%%%%%%%%
% FABRICATION
%%%%%%%%%%%%%%%%%%%%%%%%%%%%%%%%%%%%%%%%
\section{Sample fabrication} \label{sec:Fabrication}
\label{sec:Fab}

\begin{figure}[ht]
\centering
\vspace{0.5cm}
\includegraphics[width=60mm]{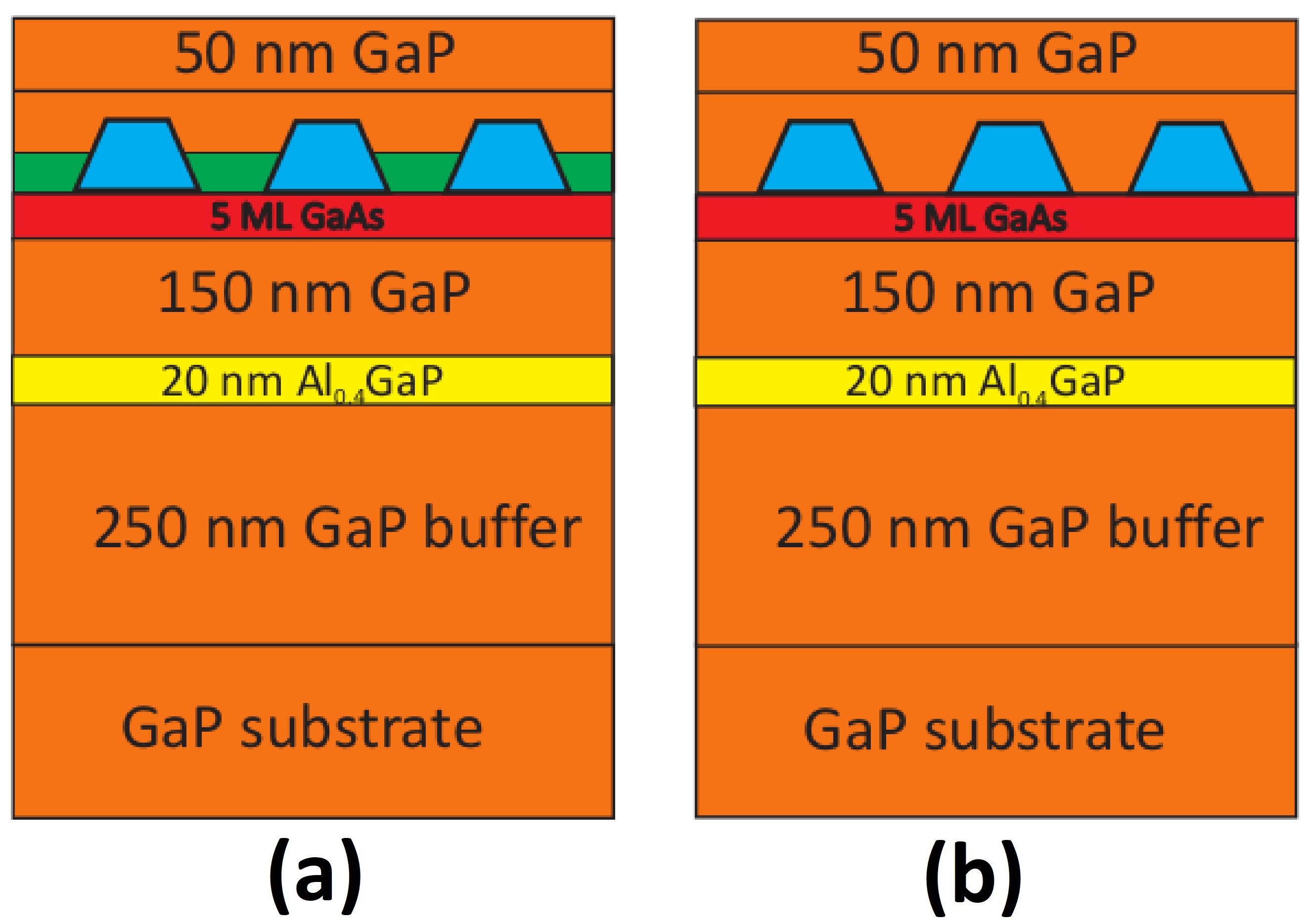}
\caption{(In,Ga)(As,Sb) QD samples studied in this work: full PL structures containing buried QDs overgrown by a thin GaSb capping layer \textbf{(a)} and reference samples with GaP capping only \textbf{(b)}.}
\label{fig:PLstructures}
\vspace{0.5cm}
\end{figure}

The samples investigated in this work are shown in Fig.~\ref{fig:PLstructures}. They were grown on GaP(001) substrates in an \textit{Aixtron Aix} 200 MOVPE reactor using H$_2$ as carrier gas, at the Technical University of Berlin. The growth commences with a 250~nm GaP buffer grown at the substrate temperature of 750$\,^\circ$C. Thereafter, a 20$\,$nm Al$_{0.4}$Ga$_{0.6}$P layer is deposited, acting as a barrier for the photogenerated carriers. This is followed by a 150~nm-thick GaP layer, before the substrate temperature is reduced to 500$\,^\circ$C. After temperature stabilization, a 5~ML GaAs interlayer (IL), a short Sb-flush, and nominal $\sim$0.51~ML $\mathrm{ In_{0.5}Ga_{0.5}Sb}$ for QD formation are deposited. The purpose of growing the GaAs interlayer is to enable the Stranski-Krastanov QD nucleation, see also Sala~\textit{et al.}~\cite{Sala2016, Sala2018, t_sala} for further details. Immediately after QD formation and at the same growth temperature, a thin GaSb layer is deposited on the QDs, with variable thicknesses of 0.4,~1,~and~1.4~ML, depending on the sample. Next, a 6~nm GaP capping layer is grown on top of the GaSb layer. The structure is then completed with a further 50~nm GaP layer grown at 620$\,^\circ$C to completely bury the QDs for PL investigations, see structure $\textbf{(a)}$ in Fig.~\ref{fig:PLstructures}. A sample without GaSb cap was investigated as reference Fig.~\ref{fig:PLstructures}~$\textbf{(b)}$. 

Finally, in order to have a better insight into the QD morphology after GaSb overgrowth, additional structures for Atomic Force Microscopy (AFM) were grown,~e.g., with free-standing QDs capped by both GaSb and the 6~nm GaP layers, and either layers, see Fig.~\ref{fig:AFMstructures}~and~\ref{fig:3_AFMs_} in Appendix~I.

%%%%%%%%%%%%%%%%%%%%%%%%%%%%%%%%%%%%%%%%
% PHOTOLUMINESCENCE
%%%%%%%%%%%%%%%%%%%%%%%%%%%%%%%%%%%%%%%%
\section{Photoluminescence investigations}
\label{sec:PL}
\begin{figure*}[htbp]
\centering
\includegraphics[width=160mm]{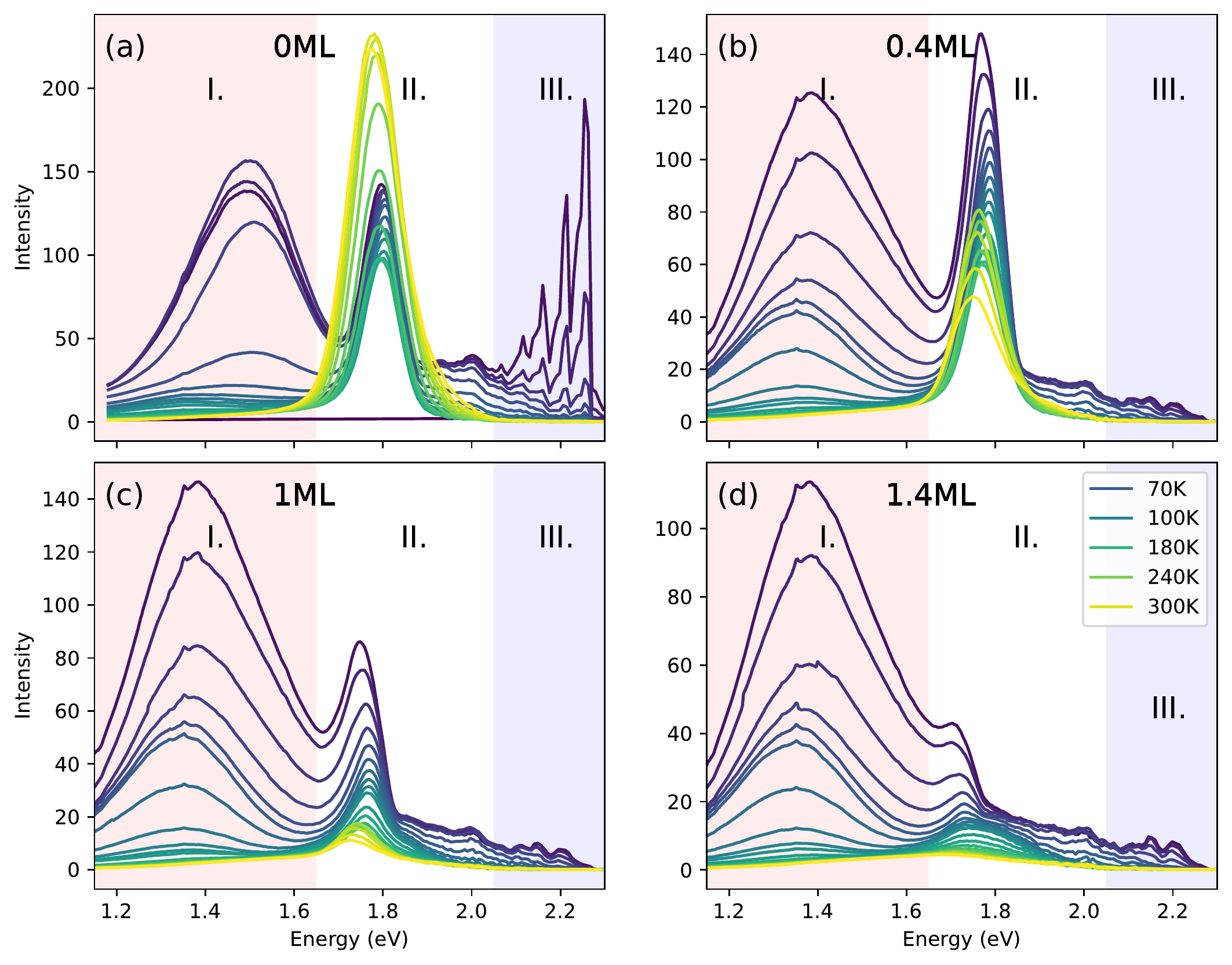}
\caption{Temperature-dependent (10-300~K) PL spectra of the four QD samples investigated in this work. From top left: (a)~QDs without GaSb cap and with (b)~0.4~ML, (c)~1~ML, and (d)~1.4~ML thick GaSb cap. The markings I.--III. refer to three distinct groups of emission,~i.e., I.~DAP (pink background), II.~QD~and~IL (white background), and~III.~neutral donor bound excitons (light blue background), see text.}
\label{fig:PL-ALL}
%\vspace{1cm}
\end{figure*}
PL measurements on ensembles of QDs were carried out on all four samples (three GaSb-capped samples and one reference without that) in the temperature range of 10 - 300~K with 10~K steps. A blue pumping laser was used with emission wavelength of 441~nm and a power density of 1~kW$\cdot$cm$^{-2}$. Fig.~\ref{fig:PL-ALL}\textbf{(a)-(d)} displays the PL spectra of the four investigated samples with varying GaSb thickness starting from no cap to 1.4~ML GaSb cap, as indicated on top of each plot. 

As general features of all spectra we observe in Fig.~\ref{fig:PL-ALL} three groups of emissions, marked by I., II., and~III. Starting from I. at the low energy side of the spectra, the broad feature arising from 1.1~eV and extending beyond 1.6~eV was previously identified by us~\cite{Steindl_2021} as donor-acceptor pair (DAP) transition~\cite{Dean_1970, Dean_1968_oxygenInGaP} localized deep in the GaP bandgap. This is followed in region II. by the luminescence peak occuring around 1.8~eV, which is of relevance for our study in this paper, and it is ascribed to the QD and IL emissions~\cite{Sala2018,Steindl_2019, Steindl_2021}.
Finally, the region III. around 2.2~eV presents transitions related to neutral donor bound excitons (D$^{0}$,$X$) which are close to the GaP bandgap at the X~${\bf k}$-point~\cite{Sala2018, Wight_1968} of the Brillouin zone (BZ). Such emission has a comparable intensity for all investigated samples and quenches fully at 300~K. Note that the energy of the GaP gap at the temperature of 10~K is 2.34~eV~\cite{PanishGaP,ioffe}. 

\begin{figure}[htbp]
\centering
\includegraphics[width=\linewidth]{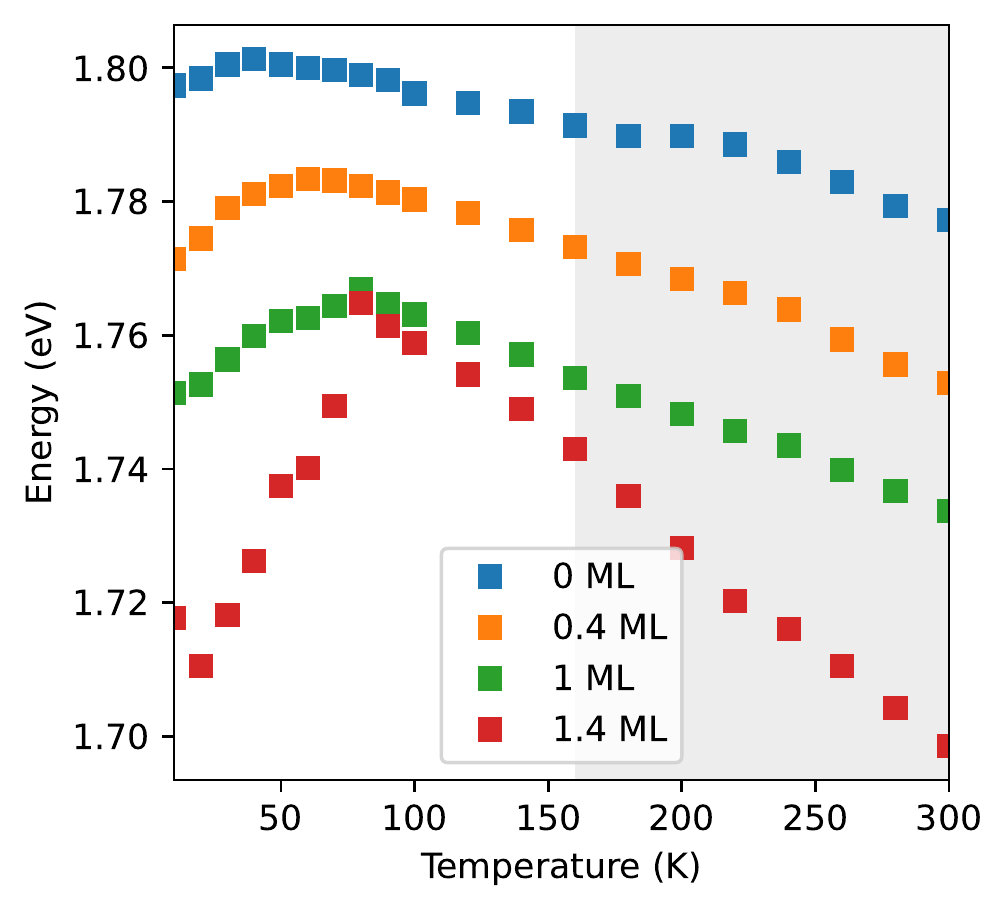}
\caption{Peak emission energies with varying temperature (10-300~K) corresponding to the QD PL band at around 1.8~eV for all samples investigated in this work. The data for temperatures higher than 160~K (grey background) were not used in evaluation by fitting in Fig.~\ref{fig:TeoPL_agreement} (see text).}
\label{fig:Varshni}
%\vspace{1cm}
\end{figure}
The QD PL band shows anomalous temperature behavior of the peak energy for different GaSb cap thicknesses, see Fig.~\ref{fig:Varshni}. There we note that (i) the peak energy of the QD PL band increases by $\sim$10~meV up to $>$60~meV for samples with GaSb cap thicknesses from 0 to 1.4~ML, respectively, when the temperature is ramped up from 10~K to $\sim$50-100~K. Moreover, (ii) the peak energy of the QD PL band does not approach zero temperature with zero gradient with respect to temperature change as expected from the Varshni dependence. Both observations (i) and (ii) seem to be contradictory to the fact that temperature dependence~\cite{Varshni} of,~e.g., electronic gap of bulk semiconductors is predominantly caused by electron-phonon interaction~\cite{ODonnell_APL1991} and approaches zero temperature with zero gradient of the energy with respect to the temperature.

Nevertheless, a similar anomalous temperature dependence, albeit of smaller magnitude, was observed in other materials before, see,~e.g., for AgGaS$_2$ alloy~\cite{YUanomalousT}, InGaAs/InP quantum wells~\cite{Skolnick_InGaAsONInPAnomalT}, or nanostructures such as InAs/GaAs QDs~\cite{ZhangInAsONGaAsAnomalT,AkibaInAsAnomalTCathodolum,JawherInAsAnomalT} or CdSe/CdS colloidal QDs~\cite{LiuCdSeCdSQDAnomalTemp}. There, statistical or rate equation models were put forward to describe that phenonenon~\cite{BansalAnomalyTtheoryModel}. In Ref.~\onlinecite{YUanomalousT} the temperature anomaly for bulk AgGaS$_2$ was explained by mixing of $p$~and~$d$ levels in valence band~\cite{Tell1972}, caused by the temperature driven lattice dilation and also due to the presence of Ag atom in the compound. However, in III-V compound semiconductors studied in this work, the $d$-levels are expected to play negligible role~\cite{Zielinsky2010,Zielinsky2012}, thus, the aforementioned argument cannot be used. On the other hand, in QD structures, such anomaly was described by rate models, which consider emission intensity transfer between various confined levels in the structure. These were,~e.g., dark and bright states~\cite{LiuCdSeCdSQDAnomalTemp}, deep-level localized states and ground~state~\cite{JawherInAsAnomalT}, wetting layer and QD states~\cite{AkibaInAsAnomalTCathodolum}, or states arising from bimodal QD size distributions~\cite{AkibaInAsAnomalTCathodolum,BansalAnomalyTtheoryModel}.

In this work, we employ the general idea of state mixing considered in the aforementioned papers in a slightly different way,~i.e., by mixing of the QD ${\bf k}$- and spatially indirect exciton states with direct ones to explain the observed temperature anomaly in our data. Hence, we first discuss our theoretical calculations in the following section.

%
%%%%%%%%%%%%%%%%%%%%%%%%%%%%%%%%%%%%%%%%
% THEORY
%%%%%%%%%%%%%%%%%%%%%%%%%%%%%%%%%%%%%%%%
%\section{Theoretical modelling}
\section{Theory model}
\label{sec:calcs}

To investigate in detail the nature of the results of our optical experiments, we performed theoretical calculations based on eight-band $\mathrm{{\bf k}}\cdot{ \mathrm{\bf p}}$ model~\cite{Stier8kp,Birner:07,Klenovsky2018_TUB,Mittelstadt2022} followed by computation of the exciton (Coulomb correlated electron-hole pair) using the configuration interaction method~\cite{Schliwa2009,Klenovsky2017,Csontosova2020,Huang2021a,Yuan2023}.

\begin{figure}[!ht]
	\begin{center}
		\begin{tabular}{c}
			\includegraphics[width=0.3\textwidth]{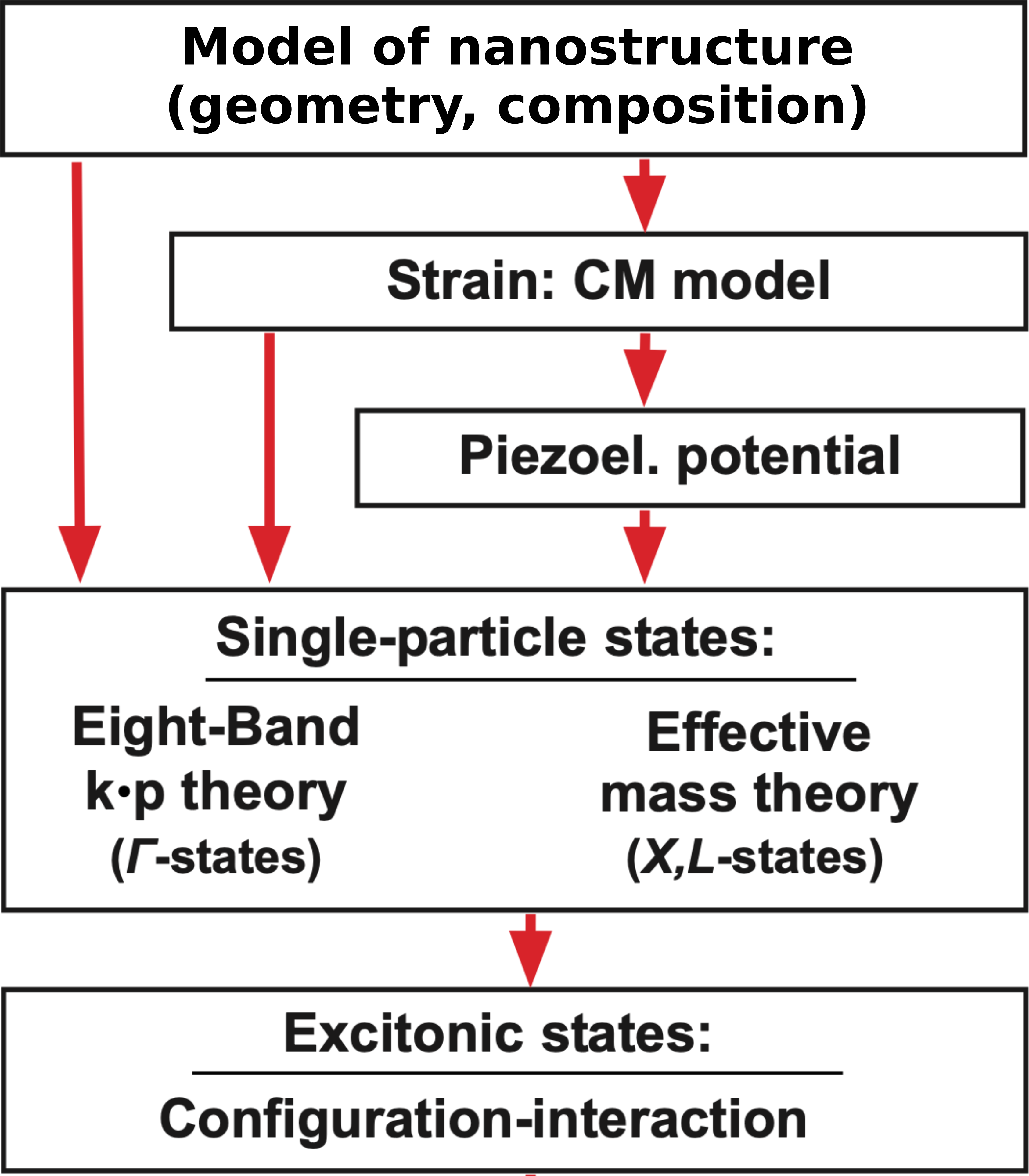}
		\end{tabular}
	\end{center}
	\caption{Flow chart of the modeling procedure applied in this work.
		\label{fig:TeoDescAndrei}}
\end{figure}
In the calculation we first implement the nanostructure model (size, shape, chemical composition), see Fig.~\ref{fig:TeoDescAndrei}. This is followed by the calculation of elastic strain by minimizing the total strain energy in the structure and subsequent evaluation of the piezoelectricity up to nonlinear terms~\cite{Bester:06,Beya-Wakata2011,Aberl2017,Klenovsky2018}. The resulting strain and polarization fields then enter either the eight-band $\mathbf{k}\cdot\mathbf{p}$ Hamiltonian for $\Gamma$ ${\bf k}$-space states, or the effective-mass Hamiltonian for L and X ${\bf k}$-space states. As a result of solving the Schr\"{o}dinger equation, we obtain electron and hole single-particle states both at the $\Gamma$- as well as at X- and L-points. The aforementioned single-particle states are then used as an input to the CI solver, which computes the energy corrections of the exciton due to the mutual electron-hole Coulomb interaction and correlation. We note that we have used two electron and two hole single-particle states as a CI basis. While larger CI basis would possibly allow for better description of the effects of the Coulomb interaction, we have refrained from that on account that the magnitude of such correction is $<2$~meV~\cite{Yuan2023,Schliwa2009},~i.e., much smaller than the overall energy resolution of excitons in our spectra here. To allow for a better reproducibility of our results in this paper, we give in detail the mathematical description of our theory methods in Appendix~II.

Using the aforementioned theory, we have computed the electronic structure of (In,Ga)(As,Sb) QDs with or without GaSb layer. Based on our previous results on this material system discussed in Refs.~\cite{Klenovsky2018_TUB,Mittelstadt2022}, XSTM measurements from Ref.~\cite{RajaNature}, and to approximately match the observed emission energy, we simulated a truncated-cone-shaped dot with base width of 15~nm and height of 3~nm. Also, for the same reasons, the QD material was chosen to be In$_{0.1}$Ga$_{0.9}$As$_{0.9}$Sb$_{0.1}$. As in our previous works, we consdiered a 5~ML GaAs layer beneath the QDs. 
\begin{figure*}[htbp]
\centering
\includegraphics[width=150mm]{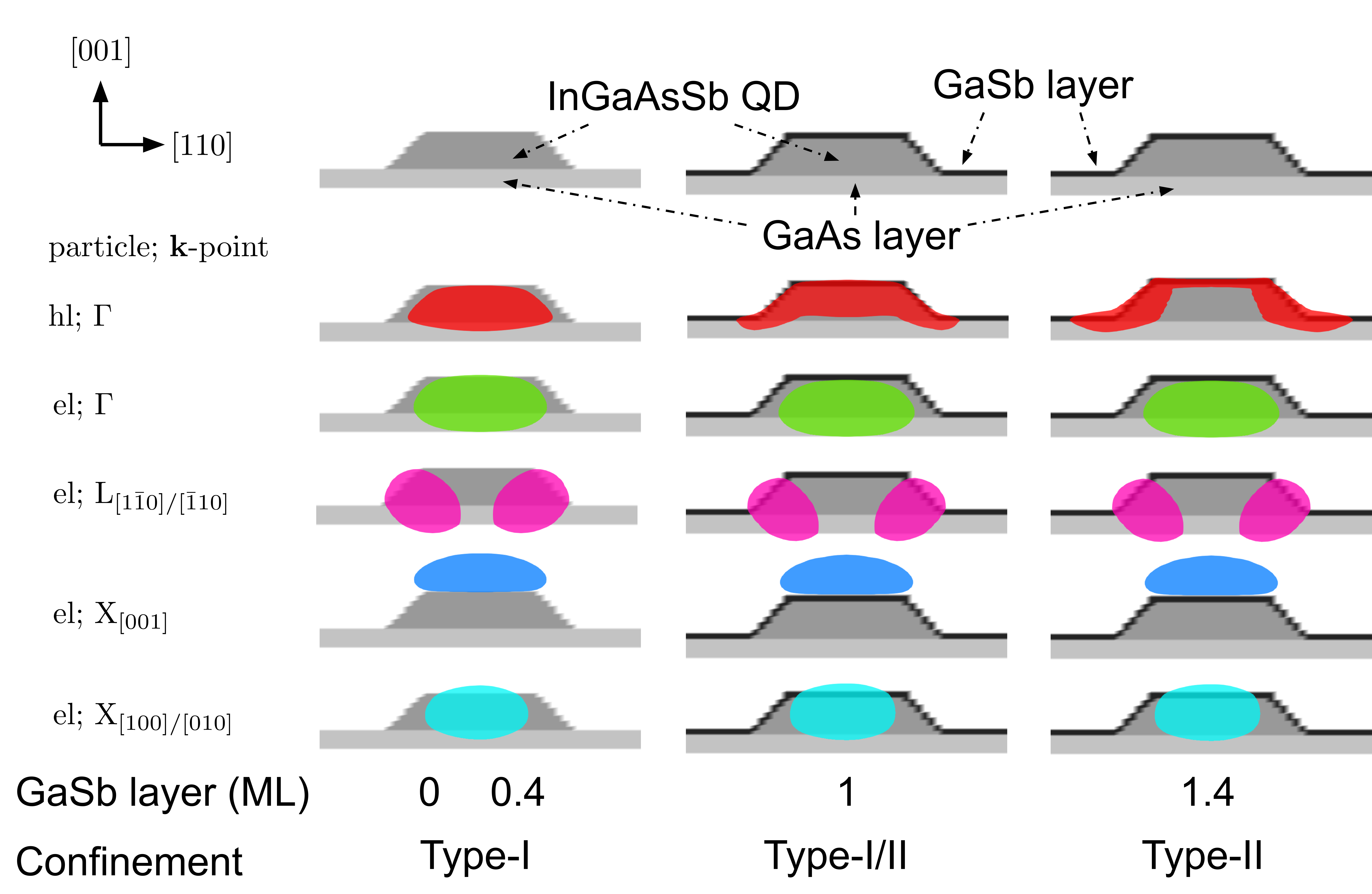}
\caption{Ground state probability densities of holes (marked as hl., red) and electrons (marked as el.). The latter is shown for $\Gamma$~(green), L~(violet) X$_{[001]}$~(dark blue), and X$_{[100]/[010]}$~(light blue) points of ${\bf k}$-space (see also main text and Ref.~\onlinecite{Klenovsky2018_TUB}). The densities are shown for GaSb layer with thicknesses of 0,~0.4,~1,~and~1.4 MLs. The corresponding identified type of confinement for $\Gamma$ electron-hole transition in case of each GaSb layer thickness is marked at the bottom of the figure. Note that 90\% of total probability density is shown.}
\vspace{1cm}
\label{fig:probab_dens}
\end{figure*}

For the QD structure described above, we show in Fig.~\ref{fig:probab_dens} the computed ground state probability densities of $\Gamma$ holes and $\Gamma$, L, and X electrons for GaSb layer thicknesses of 0~ML, 0.4~ML, 1~ML, and 1.4~ML. As already shown in Ref.~\cite{Klenovsky2018_TUB}, the L$_{[110]}$- and L$_{[1\bar{1}0]}$-electron state are predominantly elongated in [110] and [$1\bar{1}0$] crystal directions, respectively. 
We note that the L$_{[110]}$- and L$_{[1\bar{1}0]}$-electron bands are separated by less than 1~meV in our calculations,~thus, we treat in the following only the mean energies of the excitons which include L$_{[110]}$- and L$_{[1\bar{1}0]}$-electrons. For the same reason and to increase clarity of our discussion, we show in Fig.~\ref{fig:probab_dens} only one probability density for L$_{[110]}$-electron ground state. We now notice that, as the thickness of the GaSb layer above dots increases, the electron states for all discussed ${\bf k}$-points do not change appreciably. On the contrary, the $\Gamma$ hole probability density (red in Fig.~\ref{fig:probab_dens}) transforms from being fully confined in QD body (type-I band alignment) for GaSb thickness of 0~ML to being located along [110] crystal direction on the outskirts of QD, and inside of the GaSb layer for GaSb thickness of 1~ML~and~1.4~ML (type-II band alignment).

We now draw our attention to the eigenenergies of excitons formed from $\Gamma$, L, and X electrons and $\Gamma$ holes. All our structures contain 5~ML GaAs IL and we found previously~\cite{Steindl_2019,Steindl_2021} that the states of QDs and IL are energetically intermixed. We have,~thus, also computed excitons formed from $\Gamma$, L, and X electrons and $\Gamma$ holes for structure without QD and only with the 5~ML GaAs layer using our ${\bf k}\cdot{{\bf p}}$+CI toolbox, described in the aforementioned in order to compare the energies of excitons confined in IL and that in QD+IL. The calculations solely for IL were done in the same fashion as that for the QD+IL structure, only the computations for layers were performed in 1D instead of 3D.

\begin{figure}[htbp]
\centering
\includegraphics[width=\linewidth]{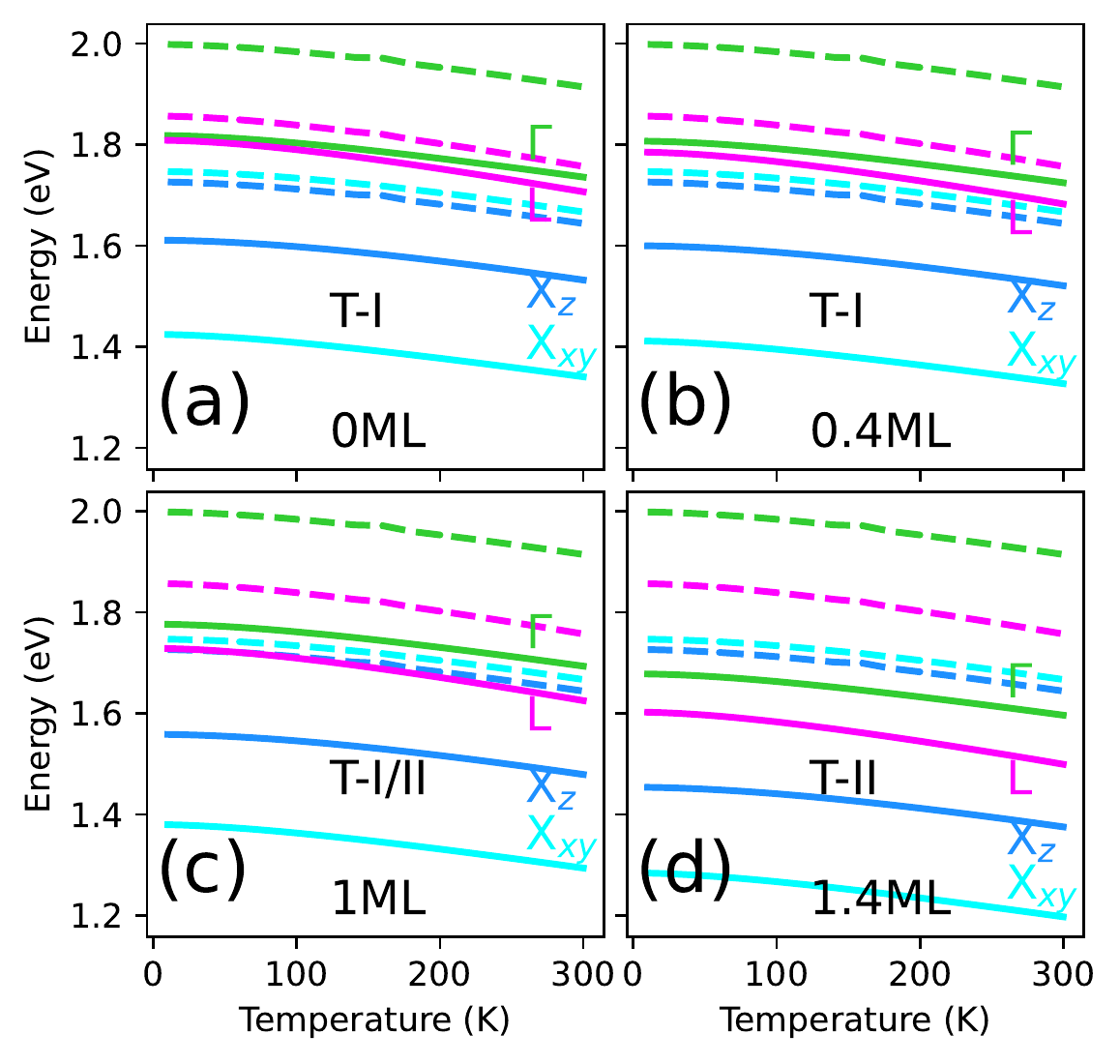}
\caption{Computed energies for states shown in Fig.~\ref{fig:probab_dens} as a function of temperature for GaSb layer thicknesses of (a) 0~ML, (b) 0.4~ML, (c) 1~ML, and (d) 1.4~ML. The energies are shown for excitons, which are composed of electrons in $\Gamma$~(green), L~(violet) X$_{[001]}$~(dark blue, marked as X$_z$), and X$_{[100]/[010]}$~(light blue, marked as X$_{xy}$) points of ${\bf k}$-space and $\Gamma$ holes. We show the energies for calculations of excitons confined in QDs and capped by GaSb by solid curves and that for GaAs interlayer (see Fig.~\ref{fig:PLstructures}) by broken curves. The markings T-I, T-II, and T-I/II indicate the type-I, type-II confinement, and transition between that. Note that the energy separation between QD excitons composed by L electrons and $\Gamma$ holes and that from $\Gamma$ electrons and $\Gamma$ holes increases with increasing GaSb layer thickness.
}
\vspace{1cm}
\label{fig:E_vs_T}
\end{figure}

In Fig.~\ref{fig:E_vs_T} we show the temperature dependencies of the excitons formed from $\Gamma$, L, and X electrons and $\Gamma$ holes for all four considered GaSb cappings. We first notice, that all energies reduce with temperature following approximately the Varshni dependence~\cite{Varshni}, see also Eq.~\eqref{eq:Varshni} in Appendix~II.
\begin{table}[!ht]
    \centering
    \begin{tabular}{c|c|c}
         {\bf k}-point & $\alpha$~(10$^{-3}$\,eV/K) & $\beta$~(K) \\
         \hline
         $\Gamma$ & 0.5405 & 204 \\
         L & 0.605 & 204 \\
         X & 0.460 & 204 \\
    \end{tabular}
    \caption{Varshni parameters~\cite{Varshni} for GaAs, taken from Nextnano++ database~\cite{Birner:07}.}
    \label{tab:VarshniGaAs}
\end{table}
As can be seen from Fig.~\ref{fig:probab_dens}, since the computed wavefunctions are located in QD and IL which both have large GaAs content in our samples, the Varshni parameters $\alpha$ and $\beta$ of the computed excitons are close to those of bulk GaAs, see Tab.~\ref{tab:VarshniGaAs}. 
\begin{figure}[htbp]
\vspace{0.5cm}
\centering
\includegraphics[width=70mm]{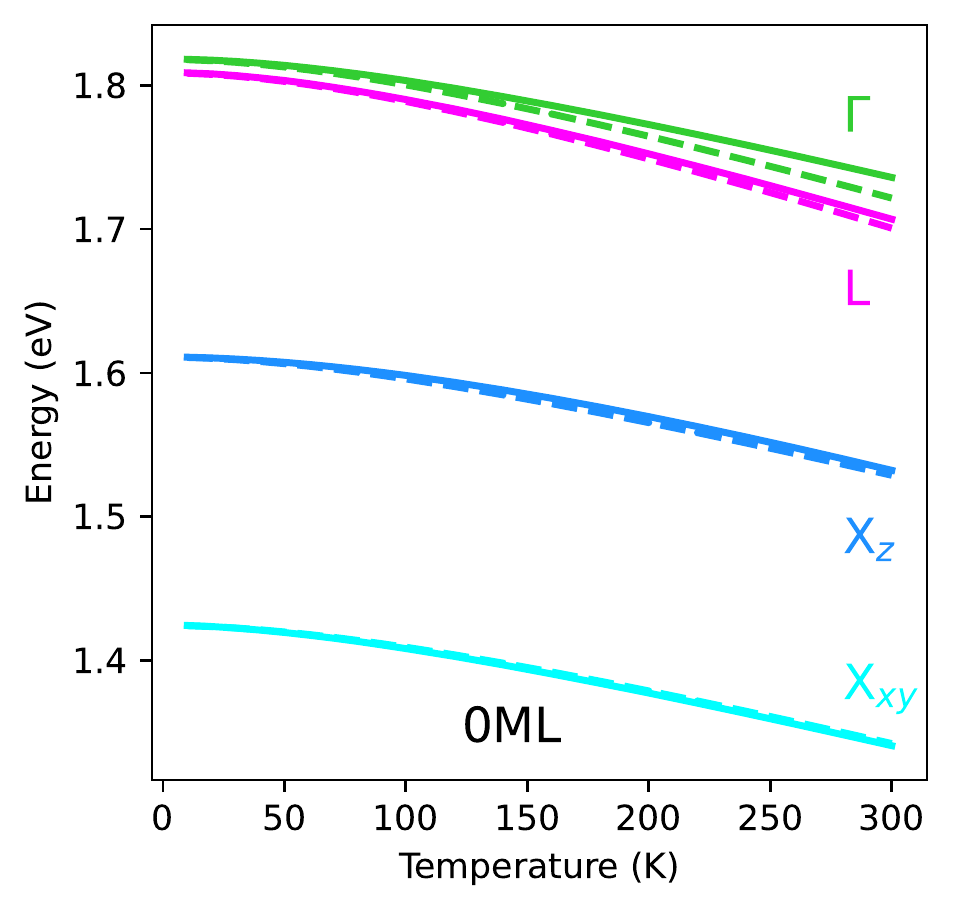}
\caption{Comparison of computed energies of excitons formed from $\Gamma$, L, and X electrons and $\Gamma$ holes of QD grown on GaAs interlayer and without GaSb capping (full curves) with corresponding Varshni~\cite{Varshni} dependence for bulk GaAs (broken curves). The data given in full curves are the same as that in Fig.~\ref{fig:E_vs_T}~(a).}
\label{fig:CIvsVarshni}
\vspace{1cm}
\end{figure}
For comparison between computed temperature dependence of exciton energies and that using Eq~\eqref{eq:Varshni} for parameters from Tab.~\ref{tab:VarshniGaAs}, see Fig.~\ref{fig:CIvsVarshni}.

We also note that in Fig.~\ref{fig:E_vs_T} the excitons confined in IL have an overall larger energy than the QD excitons. Furthermore, the binding energy due to the attractive Coulomb interaction between electrons and holes computed by our CI was found to be 20~meV and 30~meV in case of QD and IL excitons, respectively. Moreover, one can notice the mixing of IL excitons with those of QDs. All the aforementioned calculations will be employed for decomposition of the PL spectra from Fig.~\ref{fig:PL-ALL} as we will discuss in the following section. Finally, by computing the overlap integrals~\cite{Klenovsky2017,Klenovsky2018_TUB} between electron and hole states in our calculations, we have found that emission intensity of IL excitons formed from X$_{[100]/[010]}$~electrons and $\Gamma$ holes (light blue broken curves in Fig.~\ref{fig:E_vs_T}) is hundred times smaller compared to other transitions and we have, thus, omitted that from the analysis in the following section.

\section{Decomposition of photoluminescence spectra}
\label{sec:fitting}

\begin{figure*}[htbp]
\vspace{0.5cm}
\centering
\includegraphics[width=170mm]{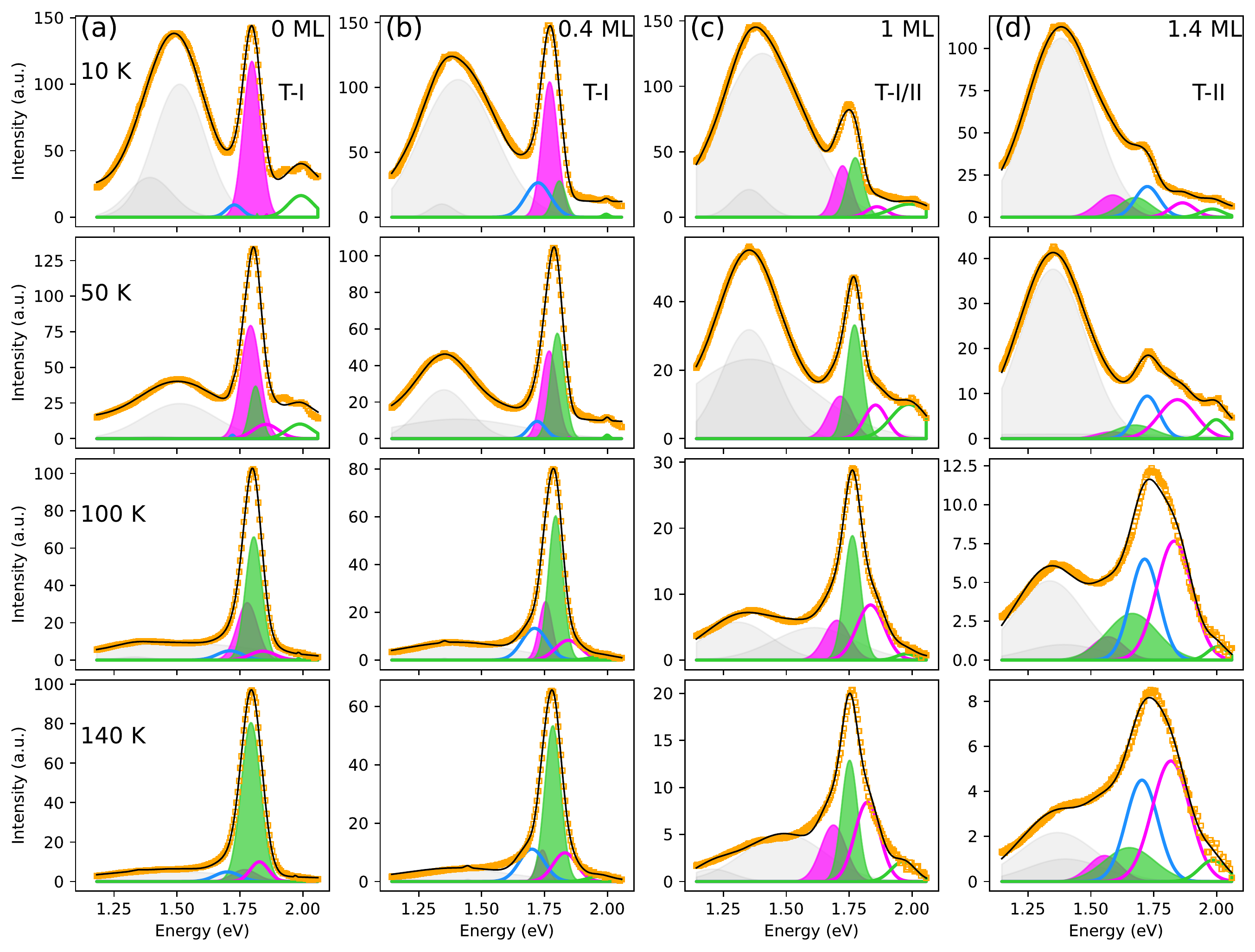}
\caption{
%Fitting results of the measured PL spectra from Fig.~\ref{fig:PL_from_QD}
%
Examples of fits (black curves) of temperature dependent PL spectra from Fig.~\ref{fig:PL-ALL} (orange open squares) by sum of seven Gaussian curves, see Eq.~\eqref{eq:Gauss}, for GaSb layer thicknesses of (a)~0, (b)~0.4~ML, (c)~1~ML, and (d)~1.4~ML. The fitted bands correspond to (i)~donor-acceptor pair (DAP) recombinations (grey), (ii)~recombination of QD exciton states (Gaussian profiles with full colored background), (iii)~GaAs interlayer exciton states (Gaussian profiles with thick colored edges and no background). The colors of the background (edges) of the Gaussian curves for QD~(IL) excitons is the same as that in Fig.~\ref{fig:E_vs_T},~i.e., excitons composed of $\Gamma$, L, and X electrons and $\Gamma$ holes are shown in green, magenta, and dark blue, respectively. The decomposition is shown for temperatures of 10,~50,~100,~and~140~K. Notice that with increase of GaSb capping thickness, the relative intensity of the interlayer bands in the spectra increases. Furthermore, as the temperature increases, the intensity of DAP recombination diminishes, particularly for temperatures $>100$~K. The vertical scale in all panels is different. The markings T-I, T-II, and T-I/II in top row of panels indicate the type-I, type-II confinement, and transition between those.
}
\vspace{0.5cm}
\label{fig:PL_fit}
\end{figure*}
We now interpret the spectra of Fig.~\ref{fig:PL-ALL} using our calculations in Fig.~\ref{fig:E_vs_T}, similarly as in our previous works~\cite{Klenovsky2012}. Our strategy is to fit the spectra from Fig.~\ref{fig:PL-ALL} by a sum of seven Gaussian functions
\begin{equation}
\label{eq:Gauss}
    I_{PL} = C+\sum_{i=1}^7I_{E_i}\exp\left\{\frac{-(\varepsilon - E_i)^2}{2\sigma_i^2}\right\},
\end{equation}
where $C$ is a constant offset of PL spectra, $\varepsilon$ is the energy axis in Fig.~\ref{fig:PL-ALL}, $\sigma_i$ the width of each Gaussian, and $I_{E_i}$ its intensity for energy $E_i$. The energies $E_i$ were set for each value of temperature to theory values from Fig.~\ref{fig:E_vs_T} with exception for X$_{[001]}$ and X$_{[100]/[010]}$ QD excitons, which could not be resolved in our spectra as those overlap with much brighter bulk DAP emission~\cite{Steindl_2019,Steindl_2021}. Thus, we interpreted the two lowest fitted energies as related to DAP bulk background transitions and we did not analyse them in detail. The remaining five energies $E_i$ corresponding to (i)~$\Gamma$-$\Gamma$ and (ii)~L-$\Gamma$ QD, and (iii)~X-$\Gamma$, (iv)~L-$\Gamma$, and (v)~$\Gamma$-$\Gamma$ IL exciton electron-hole transitions were not allowed to change by more than $\pm$1\textperthousand\,(e.g. $\pm$2 meV for energy of 1.8~eV) during the fitting routine. Example fits of PL spectra from Fig.~\ref{fig:PL-ALL} for temperatures of 10~K, 50~K, 100~K, and 140~K and all four GaSb capping thicknesses are given in Fig.~\ref{fig:PL_fit}. Note that in the fitting model for all transitions we used Gaussian profiles and not,~e.g., convolution of Lorentz with Gauss curves or more elaborate models~\cite{CHRISTEN1990,Steindl_2021}. This was motivated by the minimization of the number of fitting parameters as much as possible in order to make our fitting model more simple and,~thus, allowing more straightforward interpretation.

We see rich physics from the fits in Fig.~\ref{fig:PL_fit}. For the lowest temperatures and GaSb capping thicknesses of 0~ML and 0.4~ML related to type-I confinement, the transition around 1.8~eV is dominated by QD exciton formed from L electrons and $\Gamma$ holes. This corroborates our results from Refs.~\onlinecite{Steindl_2019,Steindl_2021}, were L QD exciton was also found to be dominant. However, as can be seen from Fig.~\ref{fig:PL_fit}~(a)~and~(b), with the sample temperature increasing, the relative intensity of the QD exciton formed from $\Gamma$ electrons and $\Gamma$ holes in the 1.8~eV band becomes more dominant. That is expected, since following Fermi-Dirac statistics of electrons, with increasing sample temperature, the probability that electrons might be thermally excited from L to $\Gamma$ QD electron state is higher.
We note, that the excitons involving L and X electrons in this work are usually called ``zero-phonon" transitions~\cite{Klenovsky2012},~i.e., phonons are not involved in the recombination of that. For the sake of completeness, transitions involving also phonons are called ``phonon replicas", but we do not consider those here as they are higher order processes with smaller probability.
Finally, we observe that the intensities of the fitted IL exciton bands are relatively smaller than those for QD exciton bands for type-I confinement in our structures. 

Furthermore, as the GaSb capping thickness increases and our system transfers from type-I to type-II confinement, the 1.8~eV band reduces in intensity. At the same time, both the relative prominence of QD excitons composed of $\Gamma$ electrons and $\Gamma$ holes increases relative to L-$\Gamma$ QD excitons. That effect can be explained by larger energy separation between $\Gamma$-$\Gamma$ and L-$\Gamma$ QD excitons when GaSb capping is increased, see Fig.~\ref{fig:E_vs_T}, thus, reducing the probability that electron is thermally excited from L QD state to that for $\Gamma$. Furthermore, with change of type of confinement the overall intensity of QD excitons reduces relative to the IL excitons, which become more visible in the whole spectra and, consequently, the relative content of those in the 1.8~eV band increases. Finally, the fitted intensity of DAP band (grey in Fig.~\ref{fig:PL_fit}) reduces with temperature both in total magnitude and relative to QD bands. For temperatures $>100$~K the DAP bands become extinct.

\begin{figure*}[htbp]
\vspace{0.5cm}
\centering
\includegraphics[width=170mm]{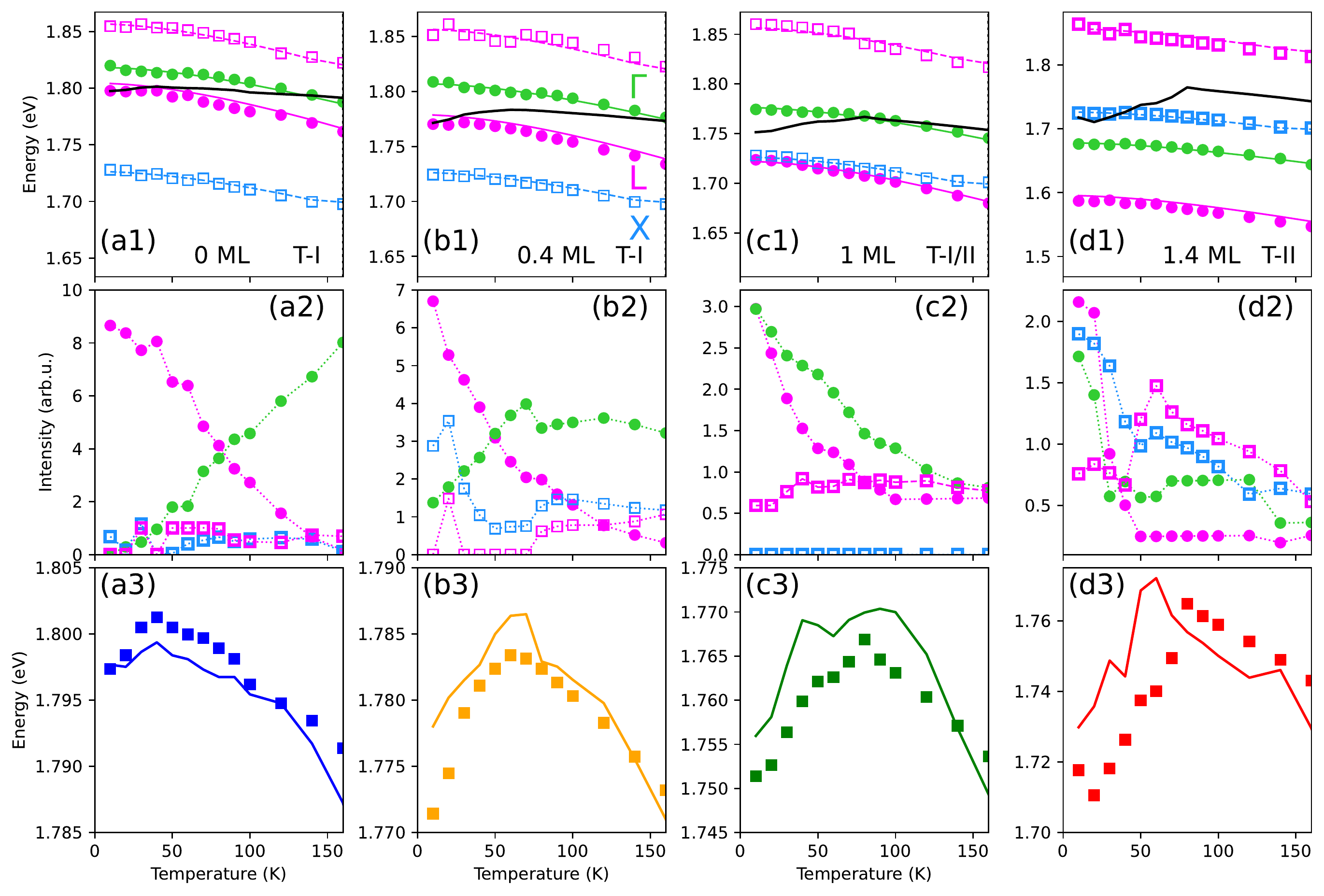}
\caption{Results of the fits by Eq.~\eqref{eq:Gauss}~(see also Fig.~\ref{fig:PL_fit}) for sample temperatures $<160$~K. The panels with labels starting with a, b, c, and d correspond to the results for GaSb capping thicknesses of 0,~0.4,~1~and~1.4~ML, respectively. In panels (a1)--(d1) we give the fitted exciton energies for QD (IL) bands as full circles (open squares). The calculated exciton energies of QD (IL), which were used as starting values for the fits (see text) are given as full (broken) curves. The excitons composed of $\Gamma$, L, X electrons and $\Gamma$ holes are given as green, magenta, and dark blue, respectively. The black curves correspond to energy values from Fig.~\ref{fig:Varshni}. In panels (a2)--(d2) we give the fitted exciton oscillator strengths (intensities), taken as integral under respective Gaussians. The colors correspond to that in panels (a1)--(d1). In panels (a3)--(d3) we show the computed,~see Eq.~\eqref{eq:TAnomal}, and measured,~see Fig.~\ref{fig:Varshni}, energy shift of the 1.8~eV band with temperature by curves and symbols, respectively. The computed shift is obtained from data in panels (a1)--(d1) and (a2)--(d2) using our model, see text. Notice that the agreement between model and theory in (a3)--(d3) is in most cases less than 5~meV. The vertical scale in all panels is different. The markings T-I, T-II, and T-I/II in top row of panels indicate the type-I, type-II confinement, and transition between those.
}
\vspace{0.5cm}
\label{fig:TeoPL_agreement}
\end{figure*}
We performed the aforementioned fitting routine for all measured PL spectra for temperatures up to 160~K. That is because for temperatures greater than 160~K, the bands in the spectra were in most cases too broadened and convoluted to be reasonably separated by fitting. The results for QD and IL excitons are shown in Fig.~\ref{fig:TeoPL_agreement}. Note that the $\Gamma$-$\Gamma$ exciton of IL was omitted from Fig.~\ref{fig:TeoPL_agreement} to not obscure the energy resolution in that figure. The fitted exciton energies are shown in Fig.~\ref{fig:TeoPL_agreement} (a1)--(d1) by full (for QD) and open (for IL) symbols. The full (for QD) and broken (for IL) curves in those panels with the same colors as the symbols represent the calculated energies from Fig.~\ref{fig:E_vs_T}. The excellent agreement between fits and theory data is because we have not allowed the values of $E_i$ in Eq.~\eqref{eq:Gauss} to change during fitting by more than $\pm$1\textperthousand\, from that calculated by theory, as we already noted earlier. Furthermore, the full black curves in Fig.~\ref{fig:TeoPL_agreement} (a1)--(d1) represent the corresponding data from Fig.~\ref{fig:Varshni} to allow for comparison of that with the computed and fitted energies. As expected, while the black curves show the anomalous temperature shift, the computed and fitted energies do not. This comparison already hints towards which energies of QD and IL excitons need to be combined to obtain the observed anomalous temperature shift of 1.8~eV band in our spectra. In the following we will elaborate on this in more detail.

In Fig.~\ref{fig:TeoPL_agreement} (a2)--(d2) we show the fitted intensities of the bands with energies given in panels (a1)--(d1). The intensities were evaluated as integral over the corresponding Gaussian bands. We see in Fig.~\ref{fig:TeoPL_agreement} (a2)~and~(b2),~i.e., for type-I regime, that the QD excitons formed from L electrons and $\Gamma$ holes switch intensity with $\Gamma$-$\Gamma$ QD excitons for temperature of $\sim 50-70$~K. As for the IL excitons, those are found to have smaller intensity for all studied temperatures in Fig.~\ref{fig:TeoPL_agreement} (a2)~and~(b2). However, the onset of type-II regime in our QDs is associated both with overall reduced intensities of QD exciton bands as well as of the ratio of intensities of L-$\Gamma$ and $\Gamma$-$\Gamma$ QD excitons. The temperature crossing of the aforementioned QD exciton species occurs in type~II for temperatures $<20$~K,~i.e., for considerably smaller temperatures than for type~I. With reduction of intensity of QD excitons, the relative prominence of the IL exciton bands in the spectra is increased. In short, the magnitude of intensities of IL exciton bands remains approximately similar in all panels (a2)--(d2), and only QD exciton intensities reduce with increasing GaSb capping thickness, as one would generally expect from the transition between type-I and type-II confinement affecting mostly QD electron states.

We now discuss an explanation of the anomalous temperature dependence shown in Fig.~\ref{fig:Varshni}. We interpret that as being due to overlapping of mutual QD and IL exciton bands in PL spectra. To model that, we take the computed energies around 1.8~eV from Fig.~\ref{fig:E_vs_T} and weight those with normalized fitted intensities from Fig.~\ref{fig:TeoPL_agreement} (a2)--(d2),~i.e.,
\begin{equation}
\label{eq:TAnomal}
    E_{\rm anomal.} = \frac{\sum_j E_jI_j}{\sum_j I_j},
\end{equation}
where $E_j$ are the {\sl computed} energies from Fig.~\ref{fig:E_vs_T}~(a)--(d)~\{also shown as curves in \ref{fig:TeoPL_agreement} (a1)--(d1)\} and $I_j$ are the fitted intensities from Fig.~\ref{fig:TeoPL_agreement} (a2)--(d2). The index $j$ goes over exciton bands close to the energy of 1.8~eV. Since the choice of summed bands and, thus, $j$ contributing to 1.8~eV band differ among samples with different GaSb capping thickness, we give that in Tab.~\ref{tab:BandsTAnomaly}. Note that the meaning of normalized oscillator strengths $I_j/(\sum_j I_j)$ is that of the occupation of the corresponding exciton state $j$.
\begin{table}[!ht]
    \centering
    \begin{tabular}{c|c}
         GaSb thickness (ML) & (${\bf k}$ el.)-(${\bf k}$ hl.) exciton \\
         \hline
         0 & $L$-$\Gamma$ QD; $\Gamma$-$\Gamma$ QD \\
         0.4 & $L$-$\Gamma$ QD; $\Gamma$-$\Gamma$ QD \\
         1 & $L$-$\Gamma$ QD; $\Gamma$-$\Gamma$ QD; $X$-$\Gamma$ IL; $L$-$\Gamma$ IL \\
         1.4 & $\Gamma$-$\Gamma$ QD; $X$-$\Gamma$ IL; $L$-$\Gamma$ IL 
    \end{tabular}
    \caption{Summed excitons in Eq.~\eqref{eq:TAnomal}. The excitons here are marked by ${\bf k}$ vector of electron (el.) and hole (hl.) and whether that originates in QD or IL.}
    \label{tab:BandsTAnomaly}
\end{table}

The results of Eq.~\eqref{eq:TAnomal} for bands in Tab.~\ref{tab:BandsTAnomaly} are shown in Fig.~\ref{fig:TeoPL_agreement} (a3)--(d3) by full curves, alongside the measured values from Fig.~\ref{fig:Varshni} given by full symbols. The agreement between the two is satisfactorily good, being less than 5~meV in most cases. It is, thus, reasonable to conclude that the temperature anomaly, which we observed, is due to mixing of different QD and also IL exciton transitions. More importantly, because we used for mixing in Eq.~\eqref{eq:TAnomal} our theory energies and not fitted ones, the temperature anomaly allowed us to show that our theory model is fully consistent with our experiment.

We finally note that our approach in Eq.~\eqref{eq:TAnomal},~i.e., the mixing of states based on their oscillator strength is similar to the approach of Refs.~\onlinecite{LiuCdSeCdSQDAnomalTemp,BansalAnomalyTtheoryModel,JawherInAsAnomalT}. However, in contrast with the aforementioned references, we do not fit the energies of the constituent transitions, but instead take without modification those computed by correlated multi-particle full CI calculations. This approach allows us to focus on the temperature emission anomaly, which we observed for different sample batch also earlier~\cite{Steindl_2019}, but did not have tools to interpret it.

\section{GaSb capping layer effect on the QD-Flash storage time}
\label{sec:qdflash}

In the previous section we showed that our theory model reproduces the experiment done on our dots very well. Now, we will use that theory model to provide expectations about the retention time of the studied QDs, when used as storage units in the QD-Flash nanomemory cell. In order to carry out the retention time measurements of the QDs, following previous designs, one has to replace the 20~nm thick Al$_{0.4}$Ga$_{0.6}$P layer of the current PL structures, see Fig.~\ref{fig:AFMstructures}, with a pure AlP layer, and embed the QD-layer in a $n^{+}-p$ diode structure~\cite{Sala2018,t_nowozinBook,t_marent,t_sala, Bonato_APL2015, Bonato2016_PSSB}. As a note, such AlP layer is used as a barrier in order to increase the hole localization energy of about $\sim$500 meV: it hinders the hole emission, but it does not affect their capture~\cite{t_nowozinBook}. 

\begin{figure}[ht]
\centering
\includegraphics[width=\linewidth]{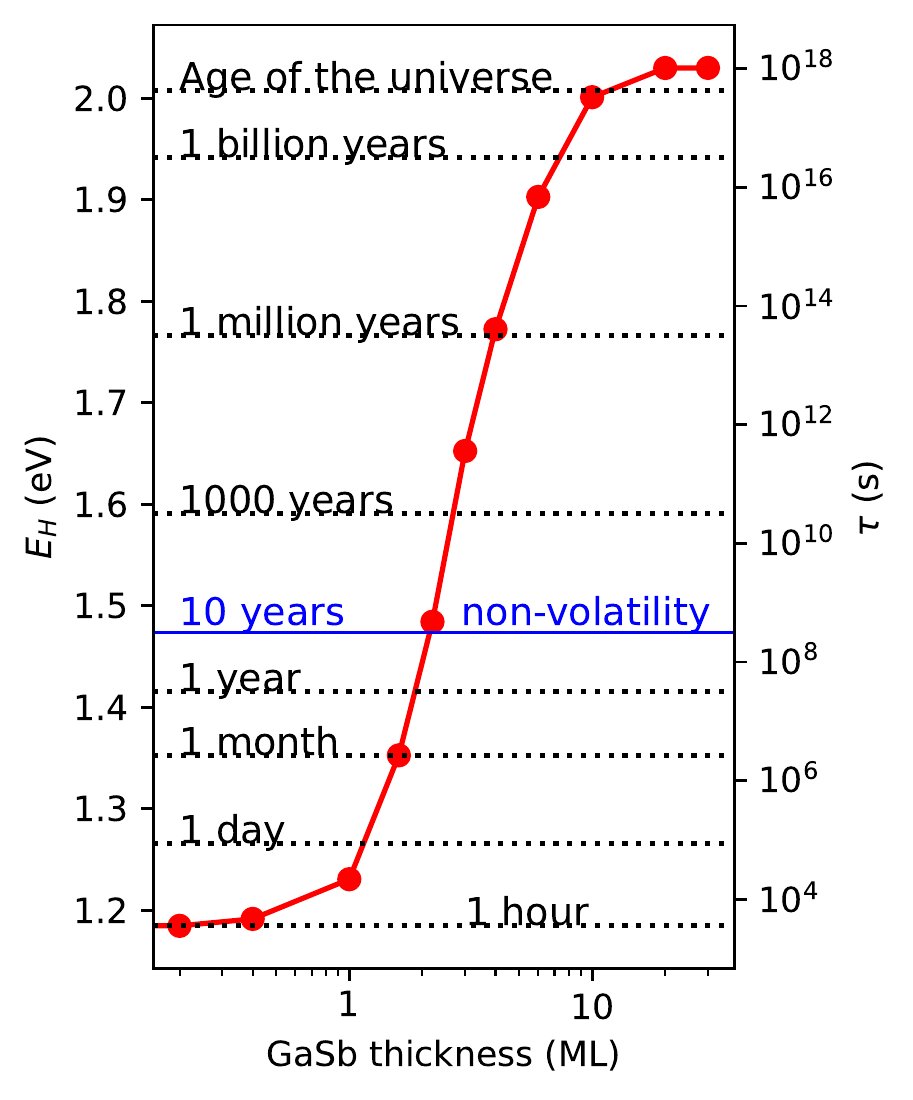}
\caption{Calculated hole binding energy ($E_{H}$) as a function of GaSb layer thickness (red balls and curves). The right vertical axis shows $E_{H}$ recomputed to storage time $\tau$ using Eq.~\eqref{eq:QDFlashStorageTime} for $T=300$~K. The horizontal dotted lines correspond to several notable time spans marked in the insets. The non-volatility limit of 10~years ~\cite{t_nowozinBook} is marked by blue horizontal line and is reached for GaSb capping thickness of 2~ML. The age of the universe of 13.8 billion years was taken from Ref.~\onlinecite{AgeUniverse}.}
\label{fig:HoleBinding}
%\vspace{1cm}
\end{figure}

Thus, here we show the prediction for the QD retention time for the QDs overgrown by a GaSb layer discussed in the aforementioned, with a 20~nm-thick pure AlP barrier layer instead of Al$_{0.4}$Ga$_{0.6}$P underneath. As shown in previous works~\cite{t_marent,t_nowozinBook,Marent2011,t_sala}, the energy $E_{H}$ can be recalculated into the storage time of QD-Flash memory units by
\begin{equation}
\label{eq:QDFlashStorageTime}
\tau=\frac{1}{\gamma\sigma_{\infty}T^2}\exp{\left(\frac{E_{H}}{k_{\rm B}T}\right)},
\end{equation}
where $T$ is the temperature, $\gamma = 1.5\times 10^{21}$~${\rm cm^{-2}K^{-2}s^{-1}}$~\cite{t_nowozinBook,t_marent}, $\sigma_{\infty}=9\times10^{-11}$~${\rm cm^2}$~\cite{Sala2018}, and $k_{\rm B}$ is the Boltzmann constant. With those parameters we computed the values of storage time $\tau$ as a function of GaSb layer thickness up to 30~ML~($\sim 10$~nm) for a temperature of $T=300$~K. The result is shown on the right vertical axis of Fig.~\ref{fig:HoleBinding}. Note that maximum of 30~ML GaSb capping was deliberatly chosen in order to study the convergence of our storage time calculations with respect to the GaSb capping thickness.
We also show the results of our calculations for the hole binding energy $E_{H}$ on left vertical axis. All the results are for In$_{0.1}$Ga$_{0.9}$As$_{0.9}$Sb$_{0.1}$ QDs grown on the 20~nm AlP layer, as outlined above.

The hole binding energy $E_{H}$ increases from $\approx1.182$~eV to $\approx2.03$~eV when the GaSb layer thickness changes from 0 to 30~ML. The minimum and the maximum $\tau$ is found to be $\tau_{\rm min}=3229$~s and $\tau_{\rm max}=33$~billions of years, respectively. We note that both the hole binding energy and the value of storage time in absence of the GaSb layer (thickness of 0~ML) in our computation match those reported {\sl experimentally} previously for a structure containing the same QDs but without any GaSb capping by Sala~\textit{et al.}~\cite{Sala2018, t_sala}. The non-volatility limit of ten years ~\cite{t_nowozinBook} is reached for GaSb capping thickness of $\sim 2$~ML. 

Remarkably, we see that adding GaSb capping over the dots leads to a huge increase of storage time by many orders of magnitude. Satisfyingly, we see from Fig.~\ref{fig:HoleBinding} that our calculation of $E_{H}$ and the associated retention time converge both for thin and thick GaSb capping layers. Note, that the time span corresponding to the age of the universe,~i.e., 13.8~billion~years~\cite{AgeUniverse} is provided solely as a time mark to highlight the potential storage time capabilities of these QDs embedded in the nanomemory device. Note, that our results in this work also confirm the previous predictions that a type-II confinement obtained by adding Sb to the QD-layer (in this case via the GaSb overgrowth), leads to a major increase in the QD retention time~\cite{Bimberg2011_SbQDFlash, t_sala}. Here, it is important to point out that such projected retention time is comparable to the one of other nanomemory device concepts currently en-route to commercialization~\cite{Ultraram}. Finally, since our QDs and nanomemory units can be fabricated using the industrially compatible large-scale MOVPE technique, they are promising candidates for a viable nanomemory device.

%%%%%%%%%%%%%%%%%%%%%%%%%%%%%%%%%%%%%%%%
% CONCLUSIONS
%%%%%%%%%%%%%%%%%%%%%%%%%%%%%%%%%%%%%%%%

\section{Conclusions}
\label{sec:conclusions}

In conclusion, we performed a detailed study of (In,Ga)(As,Sb)/GaAs quantum dots embedded in a GaP (100) matrix, which are overgrown by a thin GaSb capping layer of varied thickness, by temperature resolved photoluminescence accompanied with detailed theory.

By evaluating our photoluminescence measurements, we have found that the luminescence band corresponding to the QD emission shows anomalous temperature dependence,~i.e., increase of energy in the temperature range of 10 - $\sim$70~K, followed by energy decrease for even larger temperatures. We explained that phenomenon using ${\bf k}\cdot{\bf p}$+CI theory and Gaussian fitting of luminescence spectra using emission energies extracted from our theory. We identify, that the anomalous temperature shift is induced by mixing of different QD and interlayer excitons. This enabled us to confirm results of our electronic structure theory of the studied system and identify a change of band alignment from type-I to type-II for QDs overgrown by more than one monolayer of GaSb. Notably, we found that the ${\bf k}$-indirect electron-hole transition in type-I regime at low temperatures is optically more intense than the $\Gamma$-direct.

Finally, we provided predictions for the retention time of (In,Ga)(As,Sb)/GaAs/AlP/GaP quantum dots capped with the GaSb layer to be used as storage units in the QD-Flash nanomemory device. Strikingly, by considering a GaSb layer only 2~ML-thick, we find the projected retention time of our dots reaching the non-volatility limit of 10 years and making these QDs excellent candidates for their application in QD-Flash devices.

%%%%%%%%%%%%%%%%%%%%%%%%%%%%%%%%%%%%%%%%
% ACKNOWLEDGEMENTS
%%%%%%%%%%%%%%%%%%%%%%%%%%%%%%%%%%%%%%%%

\section{Acknowledgements}
We thank Petr Steindl for fruitful discussions and comments to our manuscript.
E.M.S. thanks the Deutsche Forschungsgemeinschaft (DFG) (Contract No. BI284/29-2). P.K. acknowledges the the projects 20IND05 QADeT, 20FUN05 SEQUME, 17FUN06 SIQUST which received funding from the EMPIR programme co-financed by the Participating States and from the European Union’s Horizon 2020 research and innovation programme. 
This work was partly funded by Institutional Subsidy for Long-Term Conceptual Development of a Research Organization granted to the Czech Metrology Institute by the Ministry of Industry and Trade.

%%%%%%%%%%%%%%%%%%%

%\bibliography{paper_TUB_PL_QDs.bib}


%\newpage 

%\vspace{15cm}

%%%%%%%%%%%%%%%%%%%%%%%%%%%%%%%%%%%%%%%%
% APPENDIX AFM
%%%%%%%%%%%%%%%%%%%%%%%%%%%%%%%%%%%%%%%%
\section*{Appendix I. Atomic Force Microscopy (AFM) investigations}
%\section{Atomic Force Microscopy (AFM) investigations}
\label{secAI:AFM}

\begin{figure*}[htbp]
\vspace{0.5cm}
\centering
\includegraphics[width=160mm]{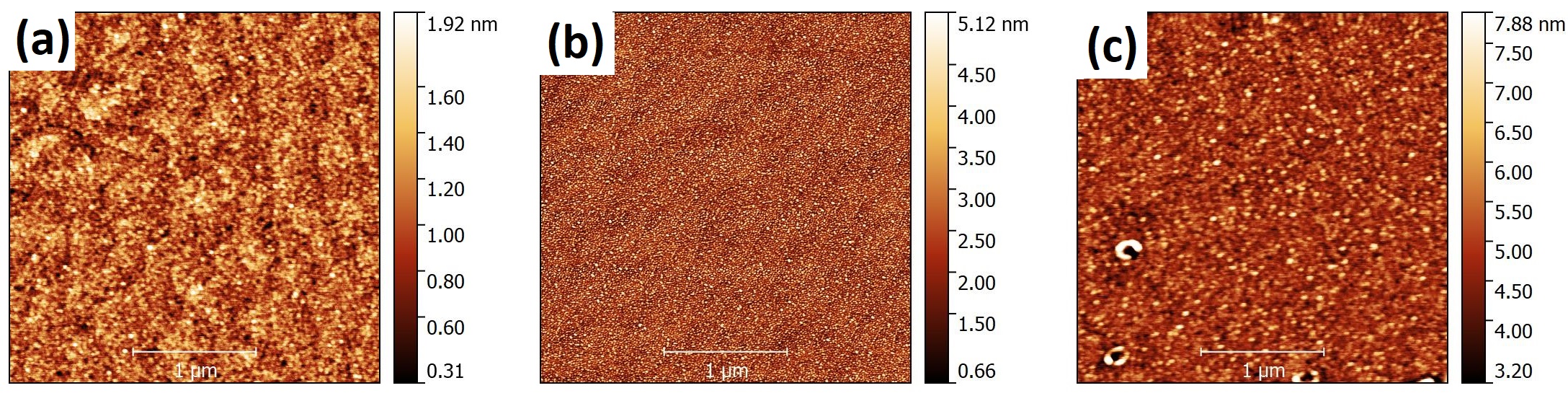}
\caption{AFM micrographs of (In,Ga)(As,Sb) QDs capped with (a) 6~nm-GaP cap only, (b) 1.4 ML GaSb layer only, and (c) both GaSb and GaP layers (scales are adjusted for better visualization of the capped QDs).} % For the layer structures of see Fig.~\ref{fig:TUstructure} $\textbf{(b)}$ and $\textbf{(c)}$.}
\label{fig:3_AFMs_}
\vspace{1cm}
\end{figure*}

In order to have a better insight on the surface morphology immediately after the QD capping, AFM measurements were carried out on the surface of (In,Ga)(As,Sb) QDs capped with only the 6~nm GaP capping layer, the 1.4~ML-thick GaSb layer, and with both GaSb and GaP layers, see layer structures in Fig.~\ref{fig:3_AFMs_}. We note that we previously carried out extensive morphological characterization of such QDs without the GaSb layer, particularly by Transmission Electron Microscopy (TEM)~\cite{Sala2016, Steindl_2019} and Cross-sectional Tunneling Microscopy (XSTM)~\cite{RajaNature}.  

\begin{figure*}[htbp]
\centering
\vspace{1cm}
\includegraphics[width=120mm]{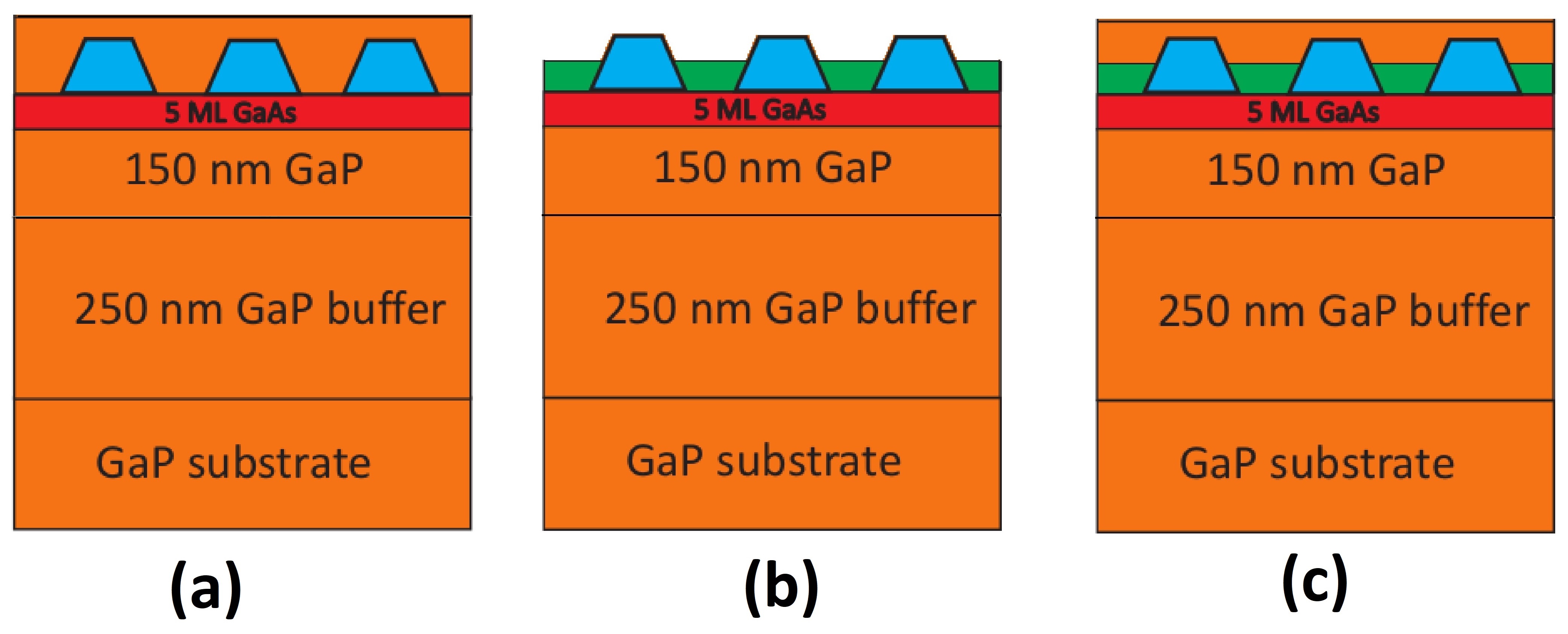}
\caption{(In,Ga)(As,Sb) QD samples studied via AFM: QDs overgrown with (a) a 6~nm GaP cap, (b) the GaSb layer and (c) both GaSb and GaP layers.}
\label{fig:AFMstructures}
\vspace{1cm}
\end{figure*}

The AFM micrographs of the samples analyzed in this work are shown in Fig.~\ref{fig:AFMstructures}~(a),~(b),~and~(c), respectively. Fig.~\ref{fig:AFMstructures}~(a) shows a very small surface modulation of about 1~nm, indicating small high-density QDs homogeneously covered with the GaP cap. Here, we point out that such QDs have an exceptionally high density of ca. 4$\cdot$10$^{11}$cm$^{-2}$, see also in Ref.~\cite{RajaNature} for more details. Fig.~\ref{fig:3_AFMs_}~(b) shows an increased modulation of ca. 2.5~nm with the largest 3D islands emerging from the background, indicating an average larger size of the QDs after deposition of the GaSb layer right after their formation. Fig.~\ref{fig:3_AFMs_}~(c) shows instead an increased surface modulation of ca. 5~nm when both the GaSb and the GaP capping layers are grown and more distinct hillocks can be observed, revealing an enlargement of the QDs upon growth of both capping layers. In addition, ring-like defects are present with a density of $\sim$10$^{7}$ cm$^{-2}$, width of 150-200~nm and height around 5-7~nm. Such structures were formed during the final GaP capping process due to a local clustering of Ga(Sb,P).

\newpage

%%%%%%%%%%%%%%%%%%%%%%%%%%%%%%%%%%%%%%%%
% APPENDIX AFM
%%%%%%%%%%%%%%%%%%%%%%%%%%%%%%%%%%%%%%%%
\section*{Appendix II. Theory methods}
%\section{Atomic Force Microscopy (AFM) investigations}
\label{secAII:Teor}
In the following sections we describe our theory model.

\subsubsection{Eight band $\mathbf{k}\!\cdot\!\mathbf{p}$ theory for $\Gamma$-states}
In eight-band ${\bf k}\cdot{\bf p}$, we consider the single-particle states as linear combination of $s$-orbital~like and $x$,~$y$,~$z$~$p$-orbital~like Bloch waves~\cite{Klenovsky2017,Csontosova2020,Mittelstadt2022} at the $\Gamma$ point of the Brillouin zone,~i.e.,
\begin{equation}
    \Psi_{a_n}(\mathbf{r}) = \sum_{\nu\in\{s,x,y,z\}\otimes \{\uparrow,\downarrow\}} \chi_{a_n,\nu}(\mathbf{r})u^{\Gamma}_{\nu}\,,
\end{equation}
where $u^{\Gamma}_{\nu}$ is the Bloch wave-function of $s$- and $p$-like conduction and valence bands at $\Gamma$ point, respectively, $\uparrow$/$\downarrow$ marks the spin, and $\chi_{a_n,\nu}$ is~the~envelope function for $a_n \in \{ e_n, h_n \}$ [$e$ ($h$) refers to electron (hole)] of the $n$-th single-particle state.
Thereafter, the following envelope-function $\mathbf{k}\cdot\mathbf{p}$ Schr\"{o}dinger equation is solved
\begin{equation}
\label{Eq:8kpSchEnvelope}
\begin{split}
\sum_{\nu\in\{s,x,y,z\}\otimes \{\uparrow,\downarrow\}}&\hat{H}_{\nu'\nu}^{\mathbf{k}\cdot\mathbf{p}}({\bf r},T)\chi_{a_n,\nu}({\bf r},T)\\
&=E^{\mathbf{k}\cdot\mathbf{p}}_n(T)\cdot \chi_{a_n,\nu'}({\bf r},T),
\end{split}
\end{equation}
where $E^{\mathbf{k}\cdot\mathbf{p}}_n(T)$ on the right side is the $n$-th single-particle eigenenergy for temperature $T$. Note that $\chi_{a_n,\nu'}$ in Eq.~\eqref{Eq:8kpSchEnvelope} is dependent also on $T$, which is due to the envelope function $\mathbf{k}\cdot\mathbf{p}$ Hamiltonian
\begin{equation}
\label{eq:HEAGamma}
\begin{split}    
    \hat{H}_{\nu'\nu}^{\mathbf{k}\cdot\mathbf{p}}({\bf r},T)&=\left[\mathcal{E}_\nu^{\Gamma}({\bf r},T)-\frac{\hbar^2{\bf \nabla}^2}{2m_e}+V_{0}({\bf r})\right]\delta_{\nu'\nu}\\
    &+\frac{{\hbar}{\bf \nabla}\cdot{\bf p}_{\nu'\nu}}{m_e}
    + \hat{H}^{\rm str}_{\nu'\nu}({\bf r})+\hat{H}^{\rm so}_{\nu'\nu}({\bf r}).
\end{split}    
\end{equation}
In Eq.~\eqref{eq:HEAGamma}, $\mathcal{E}_\nu^{\Gamma}({\bf r},T)$ is the position and temperature dependent energy of bulk $\Gamma$-point Bloch band $\nu$, $V_0({\bf r})$ is the scalar potential (e.g. due to piezoelectricity), $\hat{H}^{\rm str}_{\nu'\nu}({\bf r})$ is the Pikus-Bir Hamiltoninan introducing the effect of elastic strain,~\cite{Birner:07,t_zibold} and $\hat{H}^{\rm so}_{\nu'\nu}({\bf r})$ is the spin-orbit Hamiltonian.~\cite{Birner:07,t_zibold} Furthermore, $\hbar$ is the reduced Planck's constant, $m_e$ the free electron mass, $\delta$ the Kronecker delta, and $\nabla := \left( \frac{\partial}{\partial x}, \frac{\partial}{\partial y}, \frac{\partial}{\partial z} \right)^T$. Finally, we remark that the temperature dependence is considered in our calculations using the Varshni relation~\cite{Varshni}
\begin{equation}
\label{eq:Varshni}
    E_g(T)=E_0-\alpha T^2/(T+\beta),
\end{equation}
where $\alpha$ and $\beta$ are parameters and $E_0=\mathcal{E}^{\Gamma}({\bf r}, 0)$ is the energy projected to 0~K. The parameters $\alpha$ and $\beta$ were taken from Nextnano++ database~\cite{Birner:07} for constituent material at position ${\bf r}$.

\subsubsection{Effective mass theory for L- and X-states}
The single-particle states for L and X electrons are obtained within
the envelope function method based on effective mass approximation,~i.e.,
the following equation is solved~\cite{t_zibold} 
\begin{equation}
\hat{H}^{\mathrm{L,X}}({\pmb r},T)\chi_{n}({\pmb r},T)=E^{\rm L,X}_{n}(T)\chi_{n}({\pmb r},T),\label{Eq:singleBandEnvel-1}
\end{equation}
where $E^{\rm L,X}_n(T)$ and $\chi_{n}({\pmb r},T)$ are the $n$-th eigenenergy and the envelope
function for temperature $T$, respectively, and $\hat{H}^{\mathrm{L,X}}({\pmb r},T)$ is given by
\begin{equation}
\hat{H}^{\mathrm{L,X}}({\pmb r},T)=-\frac{\hbar^{2}}{2}{\pmb\nabla}\cdot\left(\frac{1}{\underline{m}^{*}({\pmb r})}\right){\pmb\nabla}+\mathcal{E}_{c}^{\mathrm{L,X}}({\pmb r},T)+V_{0}({\pmb r}).\label{Eq:singleBandEnvelHamiltonian-1}
\end{equation}
Here, $\mathcal{E}_{c}^{\mathrm{L,X}}({\pmb r},T)$ is the position and temperature dependent
bulk conduction band energy for L or X points, $V_{0}({\pmb r})$
is the external potential induced by,~e.g., elastic strain. The temperature dependence of $\mathcal{E}_{c}^{\mathrm{L,X}}$ is included again using Eq.~\eqref{eq:Varshni} for parameters from Nextnano++ library~\cite{Birner:07}. The effective mass parameter $\underline{m}^{*}({\pmb r})$
is given by~\cite{t_zibold} 
\begin{equation}
\underline{m}^{*}({\pmb r})=\left[m_{l}^{*}({\pmb r})-m_{t}^{*}({\pmb r})\right]\hat{\pmb k}_{0}\hat{\pmb k}_{0}^{T}+m_{t}^{*}({\pmb r}){\pmb1}_{3x3},\label{Eq:LXeffectiveMass-1}
\end{equation}
where $m_{l}^{*}({\pmb r})$ and $m_{t}^{*}({\pmb r})$ are positionally
dependent longitudinal and transversal effective masses, respectively,
$\hat{\pmb k}_{0}=\left<100\right>$ ($\hat{\pmb k}_{0}=\left<111\right>/\sqrt{3}$)
for X-point (L-point) of the Brillouin zone and ${\bf 1}_{3 \times 3}$ is
$3 \times 3$ the identity matrix.

\subsubsection{Configuration interaction for excitons}
We use single-particle states computed by the aforementioned methods as basis states for our CI. Since we consider the temperature effect already on single-particle states and that transfers into our CI calculations, we drop the dependence on $T$ in the following equations, to avoid any confusion. In CI we consider the excitonic (X) states as linear combinations of the Slater determinants
\begin{equation}
    \psi_i^{\rm X}(\mathbf{r}) = \sum_{\mathit m=1}^{n_{\rm SD}} \mathit \eta_{i,m} \left|D_m^{\rm X}\right>, \label{eq:CIwfSD}
\end{equation}
where $n_{\rm SD}$ is the number of Slater determinants $\left|D_m^{\rm X}\right>$, and $\eta_{i,m}$ is the $i$-th CI coefficient which is found along with the eigenenergy using the variational method by solving the Schr\"{o}dinger equation 
\begin{equation}
\label{CISchrEq}
\hat{H}^{\rm{X}} \psi_i^{\rm X}(\mathbf{r}) = E_i^{\rm{X}} \psi_i^{\rm X}(\mathbf{r}),
\end{equation}
where $E_i^{\rm{X}}$ is the $i$-th eigenenergy of excitonic state $\psi_i^{\rm X}(\mathbf{r})$, and~$\hat{H}^{\rm{X}}$ is the CI Hamiltonian which reads
\begin{equation}
\label{CIHamiltonian}
\hat{H}^{\rm{X}}=\hat{H}^{\rm s.p.}+\hat{V}^{\rm{X}},
\end{equation}
with $\hat{H}^{\rm s.p.}$ and $\hat{V}^{\rm{X}}$ representing the electron-hole transition energies of the non-interacting single-particle states,~i.e., either
\begin{equation}
\label{eq:HspKP-KP}
\hat{H}^{\rm s.p.}_{n,m}=E^{\mathbf{k}\cdot\mathbf{p}}_{m,e}-E^{\mathbf{k}\cdot\mathbf{p}}_{n,h},
\end{equation}
for exciton composed from $\Gamma$ electrons and $\Gamma$ holes, or
\begin{equation}
\label{eq:Hsp1b-KP}
\hat{H}^{\rm s.p.}_{n,m}=E^{L,X}_{m,e}-E^{\mathbf{k}\cdot\mathbf{p}}_{n,h},
\end{equation}
for exciton composed from L or X electrons and $\Gamma$ holes.
Further, the matrix element $\hat{V}^{\rm{X}}$ in the basis of the Slater determinants $\left|D_m^{\rm X}\right>$ is~\cite{Klenovsky2012,Klenovsky2017,Klenovsky2018_TUB,Csontosova2020}
\begin{equation}
\begin{split}
    &V^{\rm X}_{n,m}=\left<{D_n^{\rm X}}\right|\hat{J}\left|{D_m^{\rm X}}\right> = -\frac{1}{4\pi\epsilon_0} \sum_{ijkl} \iint {\rm d}\mathbf{r} {\rm d}\mathbf{r}^{\prime} \frac{e^2}{\epsilon(\mathbf{r},\mathbf{r}^{\prime})|\mathbf{r}-\mathbf{r}^{\prime}|} \\
    &\times \{ \Psi^*_i(\mathbf{r})\Psi^*_j(\mathbf{r}^{\prime})\Psi_k(\mathbf{r})\Psi_l(\mathbf{r}^{\prime}) - \Psi^*_i(\mathbf{r})\Psi^*_j(\mathbf{r}^{\prime})\Psi_l(\mathbf{r})\Psi_k(\mathbf{r}^{\prime})\}
    \\
    &= \sum_{ijkl}\left(V^{\rm X}_{ij,kl} - V^{\rm X}_{ij,lk}\right).
\end{split}
\label{eq:CoulombMatrElem}
\end{equation}
Here $\hat{J}$ marks the Coulomb operator, $e$ labels the elementary charge and $\epsilon(\mathbf{r},\mathbf{r}^{\prime})$ is the spatially dependent relative dielectric function, $\epsilon_0$ is the vacuum permittivity. Note that for $\epsilon(\mathbf{r},\mathbf{r}^{\prime})$ in Eq.~\eqref{eq:CoulombMatrElem} we use the position-dependent bulk dielectric constant. The Coulomb interaction in Eq.~\eqref{eq:CoulombMatrElem} described by $V^{\rm X}_{ij,kl}$ ($V^{\rm X}_{ij,lk}$) is called direct (exchange).

Furthermore, we note that the integrals in Eq.~\eqref{eq:CoulombMatrElem} between L and X electron states and $\Gamma$ holes are carried out only between L and X single band electron and $s$-Bloch state component of the hole eight-band ${\bf k}\cdot{\bf p}$ wavefunctions~\cite{Klenovsky2012}. That approach for excitons consisting of L and X electrons and $\Gamma$ holes is justified since, as will be discussed also later, our QDs are mostly composed of GaAs and in that material the bulk energy of L-$\Gamma$ and X-$\Gamma$ transitions between conduction and valence bands for temperature of 0~K are 1.815~eV and 1.981~eV~\cite{Birner:07}, respectively. Note, that coupling between bulk conduction and valence Bloch bands scales with the inverse of bandgap~\cite{Krapek2015,Klenovsky2018_TUB}.

Finally, the sixfold integral in Eq.~\eqref{eq:CoulombMatrElem} is evaluated using the~Green's function method.~\cite{Schliwa2009,Stier8kp,Klenovsky2017} The integral in Eq.~\eqref{eq:CoulombMatrElem} is split into solution of the Poisson's equation for one quasiparticle $a$ only, followed by a three-fold integral for quasiparticle $b$ in the electrostatic potential generated by particle $a$ and resulting from the previous step. That procedure, thus, makes the whole solution numerically more feasible and it is described by
\begin{equation}
\begin{split}
    \nabla \left[ \epsilon(\mathbf{r}) \nabla \hat{U}_{ajl}(\mathbf{r}) \right] &= \frac{4\pi e^2}{\epsilon_0}\Psi^*_{aj}(\mathbf{r})\Psi_{al}(\mathbf{r}),\\
    V^{\rm{X}}_{ij,kl} &= \int {\rm d}\mathbf{r}'\,\hat{U}_{ajl}(\mathbf{r}')\Psi^*_{bi}(\mathbf{r}')\Psi_{bk}(\mathbf{r}'),
\end{split}
\label{eq:GreenPoisson}
\end{equation}
where $a,b \in \{e,h\}$.

\end{document}